\documentclass[%
 reprint,
superscriptaddress,
 amsmath,amssymb,
 aps,pre
]{revtex4-2}

\usepackage{amssymb}
\usepackage{amsmath}
\usepackage{graphicx}
\usepackage{hyperref}
\usepackage{cleveref}
\usepackage{xcolor}
\usepackage{makecell}
\usepackage{subcaption}

\begin{document}

\preprint{APS/123-QED}
\title{Maximum entropy for dynamic processes on networks}
\author{Noam Abadi}
\author{Franco Ruzzenenti}

\affiliation{Energy and Environmental Research Institute of Groningen (ESRIG), Faculty of of Science and Engineering, University of Groningen, Groningen, 9747 AG, the Netherlands}

\begin{abstract}
Dynamic processes on networks are fundamental to understanding modern-day phenomena such as information diffusion and opinion polarization on the internet or epidemics spreading through society. However, such processes are notoriously difficult to study broadly as small changes in initial conditions, the process, or the network can lead to very different evolution trajectories. Here we apply the information-theoretic framework of maximum caliber to study the statistics of such systems analytically, focusing on processes that can be interpreted as driven by interactions between populations of different types of individuals in the network. We verify the dynamics deduced from maximum caliber by using simulations of different processes on different networks, introduce an approximation of the dynamics that significantly simplifies the problem, and show that the approximation can be used to recover well-established models of population dynamics that are typically not thought of as taking place on networks.
\end{abstract}

\maketitle

\section{Introduction}

In recent years, diffusion processes on networks have received a great deal of attention, describing, for example, epidemic and rumour propagation or opinion polarization\cite{peralta2024multidimensional,peralta2021effect,iniguez2009opinion,ruan2015kinetics, masuda2013predicting,rocha2011simulated}. Diffusion models study how entities as diverse as chemical substances, biological species, or information spread as they interact with each other and their surroundings\,\cite{Fuchs2013,Pavliotis2014}. Networks allow the capturing of arbitrary connection structures such as molecular bonds between atoms, trophic relations between species, or friendship relations in social media by representing components as nodes and connections between them as links\,\cite{zanin2013modelling,lordan2014study,laurent2015calls,toivonen2006model,adamic2000power,bianconi2005loops}. A diffusion process on a network is thereby a process spreading throughout the components of a network, so that the links determine which entities can interact at a given moment.

Moreover, the structure of the network can also evolve, either responding to the interactions or independently \,\cite{porter2016dynamical,holme2023map}. This evolution, and that of the process on the evolving structure, often exhibit random characteristics. The development of maximum entropy networks has proved a robust framework to produce network models of financial, biological and social systems, capturing their random nature statistically\,\cite{park2004statistical,garlaschelli2008maximum,squartini2011randomizingI,squartini2011randomizingII,squartini2017maximum,cimini2019statistical}. However, as maximum entropy is an equilibrium theory, it is not directly applicable to dynamic processes. 

Maximum caliber\cite{jaynes1980minimum} is an extension of maximum entropy to out-of-equilibrium statistics\cite{ghosh2020maximum, dixit2018perspective, presse2013principles, ge2012markov,jaynes1980minimum}. It can be seen as an information-theoretic analogy to the maximum entropy production principle of thermodynamics\cite{martyushev2006maximum,martyushev2021maximum}, and is therefore well-positioned to extend the success of maximum entropy networks to dynamic settings. However, its application to random dynamic networks is not easy to find in existing literature. Recently, we have applied the principle to describe random Markovian evolution of network structures\cite{abadi2024maximum}, and the same idea has been used to enhance community detection in networks in a non-Markovian setting\cite{clemente2024linking}. Here, we show that maximum caliber can also be used to capture the evolution of dynamical processes spreading through networks. We focus on processes that can be interpreted as population dynamics, understanding a population as a collection of individuals classified into different groups depending on how they interact with others, such as predatory and prey animals, or individuals infected, recovered, and susceptible to a disease. With this, we extend the application of maximum caliber from the dynamics of the network structure to processes on it.  

For this, the rest of this paper is structured as follows: in the next section, we describe dynamic processes on networks as considered for the rest of this paper, their implementation as stochastic simulations, and their analytical description through maximum caliber. In the following section, we compare results from simulations of three specific processes to those obtained by maximum caliber, with results from three different network topologies to highlight the flexibility of the method with respect to the process and the network. Finally, we discuss the general implications of our results and future work on the application of maximum caliber to networks that is not carried out in this paper.

\section{Methods}
\label{sec:methods}

The focus of this work is dynamic processes on networks where $S$ discrete states $\mathcal{S} = \{0,1,2,...,S-1\}$ spread in discrete time $t$ through nodes of a directed network with no self-loops. Each node is occupied by a single state at each time, meaning that the state of all nodes can be captured by a state vector $\vec{s}(t) \in \mathcal{S}^N$ with $N$ components, each representing the state $s_i(t) \in \mathcal{S}$ of a node $i$ at time $t$. The system evolves in two steps. The first is choosing a link of the network, which determines the pair of nodes that interact. Once the pair of nodes is chosen, the second step is to choose a new pair of states for them, while all other nodes in the network remain unaltered.

The choice of a link in the network, instead of a pair of nodes directly, reflects that the pairs of nodes that can interact are restricted to this connection structure. For example, our choice of excluding self-loops means that self-interactions never take place. Beyond this, links can be chosen in arbitrary ways from a particular connection structure. For instance, we might randomly select among all links originating from nodes in specific states, or explicitly vary how we choose a link over time. How a link in the network is chosen thus reflects our assumptions on how the underlying connection structure restricts interactions.

The states in the set $\mathcal{S}$ correspond to possible types of individuals in the system. For example, we may take $\mathcal{S} = \{0,1,2\}$ to represent prey and predators with $1$ and $2$ respectively, and $0$ for empty nodes. The choice of assigning one state to each node, namely occupying it by a maximum of one individual, makes reference to the fact that we account for the structure of connections between individuals from the smallest possible scale. If we were to consider interactions between nodes that can hold more than one individual in one or more states, for example ecosystems in different geographical locations, then we should account for the assumed interaction structure within the nodes, even if approximated for example by a regular or fully connected network. This choice influences the internal dynamics of each node as much as the connections between them, representing for example migration patterns, would influence the overall dynamics of the network.

The interpretation of node states also gives a clearer meaning to the updates of interacting pairs of nodes. For example, consider a predator prey system where a node $i$ occupied by a predator $s_i = 2$ is chosen to interact with a node $j$ occupied by a prey $s_j = 1$, so the pair of interacting states before the update is $(s_i, s_j) = (2,1)$. Imagine that after the interaction, the states are $(s_i', s_j') = (0,2)$. Because no other node in the network can change its state, the prey has been removed and the predator has taken its place leaving an empty node behind. The transition $(2,1) \rightarrow (0,2)$ can then be interpreted as a predation interaction where the predator consumes the prey. On the other hand, the transition $(2,1) \rightarrow (1,0)$ would represent the prey consuming the predator, so such a transition should be forbidden.

To do this methodically, we consider a table where each column corresponds to a pair of node states $(s,r) \in \mathcal{S}^2$ before an update and each row to a pair of states $(s',r') \in \mathcal{S}^2$ after it. This is essentially an $S^2 \times S^2$ matrix containing all potential transitions between pairs of states represented by the states $\mathcal{S}$. We must then identify ``interactions'' associated to transitions according to the interpretation of each state in $\mathcal{S}$, giving rise to what we will call the interaction table. Note that the same interaction may be associated to more than one transition, but each transition is associated a unique interaction. For example, if we consider the states $\mathcal{S} = \{0,1\}$ and interpret them as an empty node and one occupied by an individual, then a possible interaction table is given by \cref{tab:binary_interactions}. Note that forbidding transitions $00 \rightarrow 11$ and $11 \rightarrow 00$ is simply a choice capturing that only one event takes place at a time, while these transitions would represent individuals simultaneously being born spontaneously or dying. Another valid choice, for example, might be forbidding transitions associated to single and double spontaneous births but not natural or double deaths.
\begin{table}[ht!]
    \centering
    \begin{tabular}{cc|c|c|c|c}
         & t  & 00 & 01 & 10 & 11 \\
    t+1  &    &    &    &    &    \\
    \hline
    00   && No change & \makecell{Natural\\death} & \makecell{Natural\\death} & Forbidden \\
    \hline
    01   & & \makecell{Spontaneous\\birth} & No change & Movement & \makecell{Exclusive\\competition} \\ 
    \hline
    10   & & \makecell{Spontaneous\\birth} & Movement & No change & \makecell{Exclusive\\competition} \\
    \hline
    11   & & Forbidden & Reproduction & Reproduction & No change 
    \end{tabular}
    \caption{Possible interactions associated to transitions between pairs of node states, either empty ($0$) or occupied ($1$). Columns and rows represent pairs of node states before and after the transition in the corresponding cell of the table.} 
    \label{tab:binary_interactions}
\end{table}

As with the choice of the link that decides the pair of nodes that interact, the choice of a new pair of states for interacting nodes can take place in many different ways. For example, in the context of \cref{tab:binary_interactions} and given that the pair of nodes chosen to interact are in a state $(0,1)$, we may choose a new state $(0,0)$, indicating the death of the individual, more often in some nodes than others in order to capture that these locations are more harmful to the individuals. Likewise, some nodes may increase the chances of reproduction, and some links may make movement more prone to taking place. We might even use the direction of the link connecting the nodes, for example in a predator prey system to increase the chances of predation taking place when the link points from the predator to the prey but decreasing them if the link points the other way, as if indicating which notices which first.

While these choices give us the possibility to capture details of specific dynamic processes on networks, the purpose of this paper is to show how maximum caliber can be used to describe them in general. For the sake of clarity in the development of the method rather than accuracy of any particular model, we make two main simplifying assumptions, and discuss departures from them and others made in \cref{app:moments,app:links,app:ternary}. 

First, a single link is chosen at each step by drawing one uniformly from the links of a fixed network. More precisely, let $A \in \{0,1\}^{N \times N}$ be the $N \times N$ adjacency matrix of a binary directed network (where $N$ is the number of nodes), with elements at row $i$, column $j$ taking values $a_{ij} = 1$ if there is a link from $i$ to $j$ and $a_{ij} = 0$ otherwise. Let $L = \sum_{ij} a_{ij}$ be the number of links in the network. Then for any process that we consider on that network, the pair of nodes $i,j$ are chosen to interact at each step with probability
\begin{equation}
    P_G(i,j) = \frac{a_{ij}}{L} =
    \begin{cases}
        1/L & \text{if there is a link from $i$ to $j$} \\
        0 & \text{otherwise}.
    \end{cases}
    \label{eq:link_choice}
\end{equation}

Secondly, throughout the evolution of a particular process on any network, new states for any pair of interacting nodes are always chosen from the same conditional probability of a new pair of states given only the pair of states before the transition. Concretely, given the interaction table of a process, let us assign a non-negative value $v_{\alpha} \in \mathbf{R}_{\geq 0}$ to each interaction $\alpha$ in the interaction table (not to each transition), in particular with the value $v_{F} = 0$ for the forbidden interaction and $v_\text{n} = 1$ for no change, but freely chosen otherwise. Replacing the value $v_\alpha$ wherever the interaction $\alpha$ is found in the interaction table and dividing the elements of each column by their sum, we obtain  a transition matrix defining the conditional probabilities of new pairs of states given old ones. For example, from the interactions in \cref{tab:binary_interactions}, the transition matrix $B$ takes the form of \cref{tab:logistic_interactions} where each column of $B$ sums to $1$. We denote the element at the column associated to states $s,r$ and row corresponding to states $s',r'$, namely the conditional probability of the new pair of states $s',r'$ given $s,r$, as $B(s',r'|s,r)$.
\begin{table}[ht!]
    \centering
    \begin{tabular}{c|cccc}
           & 00 & 01 & 10 & 11 \\
    \hline
    00  &  n  & $d$ & $d$ &     \\
    01  & $b$ &  n  & $m$ & $c$ \\ 
    10  & $b$ & $m$ &  n  & $c$ \\
    11  &     & $r$ & $r$ &  n 
    \end{tabular}
    \caption{Transition probability matrix $B$ from the interaction table shown in \cref{tab:binary_interactions}. Empty cells correspond to forbidden transitions with probability $0$. All values are non-negative, and the sum over elements in each column is equal to $1$.}
    \label{tab:logistic_interactions}
\end{table}

\begin{figure}[h!]
    \centering
    \includegraphics[width=\linewidth]{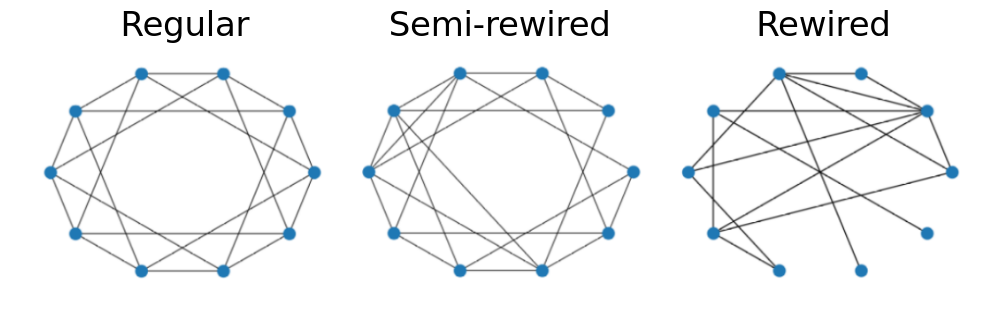}
    \caption{The three network structures for the processes considered in the following subsections. From left to right, a $10$ node regular, semi-rewired, and rewired network.}
    \label{fig:network_structures}
\end{figure}

While we fix both the distribution with which links are chosen and the conditional probability of new states for a particular process on a specific network, we consider three different processes on three different networks to test the results of maximum caliber. The three networks are shown in \cref{fig:network_structures}, and we refer to them as regular, semi-rewired and rewired. These are all directed networks, but all links are reciprocated which is why their direction is not shown. Each of these networks has the same number of nodes $N = 10$, and is drawn from a different distribution of networks with the same average number of links as the regular network $L = 4N$, but rewires a different fraction of the links. However, how these networks are obtained is not particularly relevant to the results presented here.

As for the processes considered, the first focuses on population growth limited by competition between identical individuals of a population. In this case, we choose the space of node states $\mathcal{S} = \{0,1\}$ representing an empty node and one occupied by one of these individuals respectively. The transition matrix for this process is the one given by \cref{tab:logistic_interactions}, and its interpretation is the same as given in the example that led to it.

The second process is a population with two types of individuals, prey and predators. We therefore consider the set of states $\mathcal{S} = \{0,1,2\}$ with $0$ representing an empty node and $1$ and $2$ representing prey and predators respectively. The interaction table or associated transition matrix is represented in \cref{tab:pred_prey_interactions}, which recognises the same spontaneous births, natural deaths, reproduction, movement, and competition of the previous example, but distinguishes them for prey and predators $b_\sigma$, $d_\sigma$, $r_\sigma$, $m_\sigma$ and $c_\sigma$ (with either $\sigma=1$ or $\sigma=2$) and adds a predation interaction $p$. Empty cells represent forbidden transitions, and the diagonal elements of the table, labeled n, represent no change. Note that while the reproduction of prey $r_1$ takes place across the same transitions as in competition-limited population dynamics, reproduction of the predators $r_2$ involves consuming prey.
\begin{table}[h!]
    \centering
    \begin{tabular}{c|ccccccccc}
          &  00   &  01   &  02   &  10   &  11   &  12 &  20   &  21 & 22 \\
    \hline
    00     &   n   & $d_1$ & $d_2$ & $d_1$ &      &    & $d_2$ &    &    \\
    01     & $b_1$ &   n   &      & $m_1$ & $c_1$ &    &      &    &    \\ 
    02     & $b_2$ &      &   n   &      &      &    & $m_2$ & $p$ & $c_2$ \\
    10     & $b_1$ & $m_1$ &      &   n   & $c_1$ &    &      &    &     \\
    11     &      & $r_1$ &      & $r_1$ &   n   &    &      &    &    \\
    12     &      &      &      &      &      &  n  &      &    &    \\
    20     & $b_2$ &      & $m_2$ &      &      & $p$ &   n   &    & $c_2$  \\
    21     &      &      &      &      &      &    &      & $n$ &     \\
    22     &      &      &      &      &      &$r_2$&      &$r_2$&   n  
    \end{tabular}
    \caption{Possible interactions of pairs of nodes, each either empty ($0$), occupied by prey ($1$), or occupied by a predator ($2$). Empty cells of the table are forbidden transitions.} 
    \label{tab:pred_prey_interactions}
\end{table}

The third process involves a disease spreading through a population. The states considered are $\mathcal{S} = \{0,1,2,3\}$, representing an empty node $0$, or types of individuals in this population that occupy it, namely susceptible $1$, meaning they have not yet been exposed to the disease, infected $2$, or recovered $3$ and therefore inmune from it for a time. The interactions associated to transitions between pairs of these states are given in \cref{tab:SIR_interactions}. These involve natural births, deaths and movement of each type of individual $b_\sigma$, $d_\sigma$ and $m_\sigma$ respectively (for $\sigma = 1$, $2$ or $3$). They also account for the reproduction of individuals producing another of their own type $r_\sigma$, and also the possibility that infected and recovered individuals reproduce as susceptible ones $r_{\sigma1}$. Additionally, interactions specific to the spreading of the disease are contagion $c$, healing from the disease $h$, loss of immunity after recovering $l$, and switching the location of an infected individual $s$.
\begin{table}[h!]
    \centering
    \begin{tabular}{c|cccccccccccccccc}
          &  00   &  01   &    02   &  03   &  10   &  11   &  12  &  13  &  20    &  21  &  22  &  23  &  30    &  31  &  32  &  33  \\
    \hline
    00    &   n   & $d_1$ &  $d_2$  & $d_3$ & $d_1$ &       &      &      & $d_2$  &      &      &      & $d_3$  &      &      &      \\
    01    &       &   n   &         & $l$   & $m_1$ &       &      &      &        &      &      &      &        &      &      &      \\
    02     &       &       &    n    &       &       &       &      &      &  $m_2$ &      &      &      &        &      &      &      \\
    03     &       &       &   $h$   &   n   &       &       &      &      &        &      &      &      & $m_3$  &      &      &      \\
    10          & $m_1$ &         &       &   n   &       &      &      &        &      &      &      &  $l$   &      &      &      \\
    11     &       & $r_1$ &         &       & $r_1$ &   n   &      &      &        &      &      &      &        &      &      &      \\
    12     &       &       &$r_{21}$&        &       &       &  n   &      &        &      &      &      &        &      &      &      \\
    13     &       &       &        &$r_{31}$&       &       &      &  n   &        &      &      &      &        & $s$  &      &      \\
    20     &       &       &  $m_2$ &        &       &       &      &      &   n    &      &      &      &        &      &      &      \\
    21    &       &       &        &        &       &       &      &      &$r_{21}$&  n   &      &      &        &      &      &      \\
    22     &       &       &  $r_2$ &        &       &       &  $c$ &      & $r_2$  & $c$  &  n   &      &        &      &      &      \\
    23     &       &       &        &        &       &       &      &      &        &      &      &  n   &        &      &      &      \\
    30     &       &       &        & $m_3$  &       &       &      &      &  $h$   &      &      &      &   n    &      &      &      \\
    31     &       &       &        &        &       &       &      & $s$  &        &      &      &      &$r_{31}$&  n   &      &      \\
    32     &       &       &        &        &       &       &      &      &        &      &      &      &        &      &  n   &      \\
    33     &       &       &        & $r_3$  &       &       &      &      &        &      &      &      & $r_3$  &      &      &  n   
    \end{tabular}
    %}
    \caption{Interactions of two nodes, each either empty ($0$), susceptible ($1$), infected ($2$) or recovered ($3$). Empty cells of the table are forbidden transitions.}
    \label{tab:SIR_interactions}
\end{table}

Dynamic processes on networks such as those described are often implemented through numerical simulations. However, the random nature of the processes implies that multiple repetitions of the same process on the same network must be computationally evaluated to measure its statistical properties. The amount of repetitions required increases with the number of links in the network and the number of steps over which the process is carried out. This can quickly become prohibitively high, for example in large networks. Part of the motivation of developing the analytical method of maximum caliber for such systems is driven by this high cost of simulations. However, we still implement simulations of the processes to serve as ground truths for the statistics of the system against which to compare the results of maximum caliber, driving our choice of only $N = 10$ nodes for the networks considered. 

\subsection{Simulations}

Simulations of the processes described were implemented in Python to evaluate the evolution of node states numerically and compare the statistics of their results to maximum caliber. For each process, we first chose a single initial condition $\vec{s}(0)$ and the values of the transition matrix, both selected randomly. The initial condition and transition matrix remain the same as long as the process considered is, regardless of the network on which it takes place, the time elapsed of the process or repetitions to obtain statistical results.

For each network, and starting from the initial condition chosen for each process, a single update of node states is carried out in two steps. First, a pair of nodes is drawn uniformly among those connected by the links of the network, i.e. according to the probability in \cref{eq:link_choice}. Second, a new pair of states is chosen according to the probabilities in the column of the transition matrix associated with the initial pair of states. This update is iteratively repeated $T=300$ times defining a single realisation of the evolution of the states of the process on the network from the chosen initial condition. 

For each process and network considered, $R = 100000$ realisations were carried out to measure statistical properties. We consider all $9$ possible combinations of the $3$ networks and $3$ processes described. The central statistical property of interest in each of these is the probability $\rho_i(s)$ of a particular node $i$ being found in the state $s$ and how this probability evolves over time. This object can be viewed as a non-negative valued matrix of $N$ rows, corresponding to nodes, and $S$ columns, associated with states, that evolves in time. We do not make time dependence explicit in the notation of probabilities to simplify it. To measure $\rho_i(s)$ at a particular time, consider the state vector $\vec{s^n}(t)$ in the $n$-th realisation of a process at time $t$. From it, we construct an $N \times S$ binary matrix with the element at row $i$, corresponding to a node, and column $s$, associated with a state, taking values $C^n_i(s) = 1$ if $s^n_i(t) =s$ and $C^n_i(s) = 0$ otherwise. Note that this is essentially the Kronecker delta $C^n_i(s) = \delta(s^n_i(t), s)$. We then measure the probability of node $i$ being in the state $s$ as the average amount of times this happens in the realisations,
\begin{equation}
    \rho_i(s) = \frac{1}{R} \sum_n C^n_i(s) \, .
    \label{eq:state_probability_estimation}
\end{equation}

One of the main results of maximum caliber is that the evolution of probability of finding nodes in specific states $\rho_i(s)$ is related to the probability $\rho_{ij}(s,r)$ of finding a pair of nodes $i,j$, regardless of whether they are linked or not, in states $s,r$ respectively at each time. The probabilities can be pictured as $S^2$ matrices, each associated to a pair of states $s,r$, with $N \times N$ non-negative valued elements, corresponding to pairs $i,j$ of nodes. To verify the relation obtained from maximum caliber, we also measure this joint probability of pairs of node states at each time from simulations. For this, we construct $S^2$ binary matrices, each associated to a pair of states, of size $N \times N$, with elements corresponding to pairs of nodes, from the state vector $\vec{s^n}(t)$ of each realisation at a specific time. The matrix corresponding to the pair of states $s,r$ at row $i$, column $j$ takes values $C^n_{ij}(s,r) = 1$ if both $s^n_i(t) = s$ and $s^n_j(t) = r$, or $C^n_{ij}(s,r) = 0$ otherwise. This can be expressed as a product of Kronecker deltas $C^n_{ij}(s,r) = \delta(s^n_i(t), s) \delta(s^n_j(t), r)$. The joint probability of nodes $i$ and $j$ in states $s$ and $r$ respectively is then estimated as
\begin{equation}
    \rho_{ij}(s,r) = \frac{1}{R} \sum_n C^n_{ij}(s,r) \, .
    \label{eq:state_pairs_probability_estimation}
\end{equation}

The connection between the joint probability of pairs of node states and the evolution of the probability of states of each node is essentially made through the probability $\rho_i(s,s')$ of finding, at a specific node $i$, a state $s$ at one time and another $s'$ at the next. This object can be interpreted as a non-negative matrix of $N$ rows, each associated to a node $i$, and $S^2$ columns, corresponding to pairs of consecutive states $s,s'$. Note that while \cref{eq:state_probability_estimation,eq:state_pairs_probability_estimation} each apply to a single time in the evolution of the process, the probability of successive node states describes a step from each time to the next. To measure it from simulations, we consider the state vector of a particular realisation at two successive times, $\vec{s^n}(t)$ and $\vec{s^n}(t+1)$. We then construct an $N \times S^2$ binary matrix with elements $C^n_i(s,s') = 1$ if both $s^n_i(t) = s$ and $\vec{s^n}_i(t+1) = s'$, or $C^n_i(s,s') = 0$ otherwise. This is also a product of deltas, $C^n_i(s,s') = \delta(s^n_i(t), s) \delta(s^n_i(t+1), s')$. The joint probability of consecutive states of nodes is then obtained from all realisations as
\begin{equation}
    \rho_{i}(s,s') = \frac{1}{R} \sum_n C^n_{i}(s,s') \, .
    \label{eq:consecutive_states_probability_estimation}
\end{equation}

\subsection{Maximum caliber}
\label{sec:methods:maxcal}

We now turn to how maximum caliber captures the probability of states in each node and its evolution over time. Maximum caliber establishes how to calculate distributions of dynamic system trajectories. For a discrete time process, these trajectories are essentially the sequence of states $X_T = (X(0),X(1),...,X(T))$ that capture the evolution up to time $T$ of the system. To calculate the probability $\rho(X_T)$ of each of these trajectories, maximum caliber asserts that we should first specify a series of average values of the distribution, also known as constraints
\begin{equation}
    a_n = \sum_{X_T} F_n(X_T) \rho(X_T) \, .
    \label{eq:trajectory_constraints}
\end{equation}
Then, out of all distributions that satisfy these constraints, we should choose the one that also maximises the Shannon entropy of the trajectory distribution
\begin{equation}
    S = -\sum_{X_T} \rho(X_T) \ln \left( \rho(X_T) \right) \, .
\end{equation}
The distribution that satisfies these properties can be shown\cite{presse2013principles} to take the functional form
\begin{equation}
    \rho(X_T) \propto e^{ -\sum_n \lambda_n F_n(X_T) }
    \label{eq:maxcal_distribution}
\end{equation}
where $\lambda_n$ are scalar values known as Lagrange multipliers. The distribution itself is then defined by specifying Lagrange multipliers in this functional form that guarantee that, when normalised, the distribution in \cref{eq:maxcal_distribution} takes the numerical values of the average specified by the left-hand side of \cref{eq:trajectory_constraints}. Because this numerical value only changes the values of Lagrange multipliers, it is also common to define only the functions constrained, and study the properties of the distribution for different $\lambda_n$.

Moreover, \cref{eq:maxcal_distribution} can be used to calculate conditional probabilities describing how a new state is added to a trajectory at each time in the evolution. For this, the probability of $T$-step trajectories is first interpreted as the joint probability of a $T-1$-step trajectory $X_{T-1}$ and final state $X = X(T)$, $\rho(X_T = (X_{T-1}, X)) = \rho(X_{T-1}, X)$. We can make this dependence explicit in constraints as well, writing $F_n(X_{T-1}, X):= F_n(X_T = (X_{T-1}, X))$. 

If we then marginalise the joint distribution over the new states $X$, we should obtain the functional form of the distribution of trajectories one step shorter,
\begin{equation}
    \begin{aligned}
        \rho(X_{T-1}) &= \sum_X \rho(X_{T-1}, X) \propto Z(X_{T-1}) \\
        Z(X_{T-1}) :&= \sum_x e^{- \sum_n \lambda_n F_n(X_{T-1}, X)} \, .
    \end{aligned}
\end{equation}
Since both these distributions have the same proportionality factor, their conditional probability will be
\begin{equation}
    \rho(X | X_{T-1})= \frac{\rho(X_{T-1}, X)}{\rho(X_{T-1})} = \frac{e^{- \sum_n \lambda_n F_n(X_{T-1}, X)}}{Z(X_{T-1})} \, .
    \label{eq:conditional_probability}
\end{equation}
The function $Z(X_{T-1})$ then acts as a normalisation factor that guarantees that the conditional probability obeys measure conservation for any trajectory before the state is added, that is $\sum_X \rho(X|X_{T-1}) = 1 ~~ \forall ~ X_{T-1}$. 

While we have imagined taking a $T$-step distribution and related it to that of $T-1$ steps, with the conditional probabilities of new states we see how to update the probabilities of $T-1$-step trajectories to those of $T$ steps and calculate the probability $\rho_T(X)$ of the state $X$ at time $T$,
\begin{equation}
    \begin{aligned}
        \rho(X_T = (X_{T-1}, X)) = &\rho(X | X_{T-1}) \rho(X_{T-1})\\
        \rho_T(X) = \sum_{X_{T-1}} &\rho(X | X_{T-1}) \rho(X_{T-1}) \, .
    \end{aligned}
    \label{eq:update}
\end{equation}
The fact that the conditional probability \cref{eq:conditional_probability} conserves measure then guarantees that if the $T-1$-step distribution is normalised, so will the $T$-step one. Thus, from any initial distribution $\rho_o(X)$, which can be interpreted as the probability $\rho(X_o) = \rho_o(X)$ of $0$-step trajectories $X_o = (X)$, \cref{eq:update} can be used to recursively update the trajectory distribution through conditional probabilities defined by a choice of constraints according to \cref{eq:conditional_probability}. 

However, the set of all possible $T$-step trajectories (trajectories at time $T$) is a space of size $\mathcal{X}^T$ where $\mathcal{X}$ is the number of possible states at a particular time. Implementing \cref{eq:update} directly, for example to integrate the trajectory distribution numerically, can then quickly become as demanding as simulations. Nevertheless, the central object of interest is the distribution of states at each time $\rho_T(X)$, which requires the entire trajectory distribution $\rho(X_T)$ to be calculated through maximum caliber only in principle. If the expression for $\rho_T(X)$ in \cref{eq:update} can be simplified, for example due to chosen constraints, then the entire space of trajectories may not be needed. 

In particular, note that if constraints $F_n(X_T) = F_n(X',X)$ on $T$-step trajectories $X_T$ depend only on the last two states in it $X' = X(T-1)$ and $X = X(T)$, then the conditional probability results in $\rho(X|X_{T-1}) = \rho(X|X')$. In this case, we can write
\begin{equation}
\begin{aligned}
    \rho_T(X) =& \sum_{X_{T-1}} \rho(X|X_{T-1}) \rho(X_{T-1}) \\
    =& \sum_{X'} \rho(X|X') \sum_{X_{T-2}} \rho(X'|X_{T-2}) \rho(X_{T-2}) \\
    =& \sum_{X'} \rho(X|X') \rho_{T-1}(X') \, ,
\end{aligned}
\end{equation}
where we first separate the sum over $T-1$-step trajectories into final states $X'$ and $T-2$-step trajectories $X_{T-2}$, and express $\rho(X_{T-1}) = \rho(X',X_{T-2})$. We then factor $\rho(X|X')$ out of the sum over $X_{T-2}$, and recognise the result of this sum as the probability of states $X'$ at time $T-1$. Thus, the entire trajectory distribution is not needed, only the probabilities of individual states at the previous time, meaning that the process is Markovian. Likewise, if the constraints depend only on the last $\tau$ states of the trajectory, we can divide the sum over $T-1$-step trajectories into these last $\tau-1$ steps and the former $T-1-\tau$ ones, reducing the description to distributions of $\tau-1$ steps $\rho_T(X) = \sum_{X_{\tau-1}} \rho(X|X_{\tau-1})\rho(X_{\tau-1})$.

To apply maximum caliber to dynamic processes on networks considered here, note that the variables needed to describe the evolution at each time consist of the state vector and chosen link $X(t) = (\vec{s}(t), ij(t))$. In principle, trajectories would then consist of sequences of such pairs. However, because the two variables at each time are also chosen one after the other, we can consider trajectories as consisting of alternating sequences of state vectors and links, that is $X_T = (\vec{s}(0), ij(1), \vec{s}(2), ..., \vec{s}(T-2), ij(T-1), \vec{s}(T))$ (assuming even $T$ is even). Note that in this perspective, two steps go between two consecutive state vectors, meaning that one step in the simulations corresponds to two in maximum caliber. The history of this trajectory is $X_{T-1} = (\vec{s}(0), ..., \vec{s}(T-1), ij(T-1))$ and its last state is $X = \vec{s}(T)$. 

To specify the transition probabilities in \cref{eq:conditional_probability}, the constraint functions $F_n(X_{T-1},X)$ must be defined. To obtain transitions corresponding to dynamic processes on networks, we first define what we will refer to as selector functions of each transition. These functions ``detect'' when a particular interaction has taken place from two pairs of states. For example, the selector function for the spontaneous birth interaction in the competition-limited system with interactions in \cref{tab:logistic_interactions} is
\begin{equation}
    f_b(s,r,s',r') = \begin{cases}
        1 & \text{if $(s,r) = (0,0)$ and $(s',r') = (1,0)$} \\
        1 & \text{if $(s,r) = (0,0)$ and $(s',r') = (0,1)$} \\
        0 & \text{otherwise.}
    \end{cases}
    \label{eq:birth_selector}
\end{equation}
For the movement interaction, the selector function is
\begin{equation}
    f_m(s,r,s,r) = \begin{cases}
        1 & \text{if $(s,r) = (0,1)$ and $(s',r') = (1,0)$} \\
        1 & \text{if $(s,r) = (1,0)$ and $(s',r') = (0,1)$} \\
        0 & \text{otherwise.}
    \end{cases}
    \label{eq:movement_selector}
\end{equation}
And in general, for each interaction $\alpha$ in a process, the selector function is
\begin{equation}
    f_\alpha(s,r,s',r') = \begin{cases}
        1 & \text{if $s,r \rightarrow s',r'$ is a transition} \\
          & ~~~~\text{of the interaction $\alpha$,} \\
        0 & \text{otherwise.}
    \end{cases}
\end{equation}
Note that there is a selector function for forbidden transitions which takes a value of $1$ if the transition is a forbidden one, and $0$ otherwise. However, while there is no selector function for the identity transition of pairs of states $s,r \rightarrow s,r$, we do define the identity selector function for single states, which detects when the state has remained unchanged. This is essentially the Kronecker delta $f_I(s,s') = \delta(s,s')$.

With the selector functions, we can now construct the constraint functions of the process. First, to simplify notation, let $\vec{s}(T-2) = \vec{s}$, $ij(T-1) = ij$ and $\vec{s}(T) = \vec{s}'$. For each interaction $\alpha$ of a process, we define a constraint function $F_n(X_{T-1}, X)$ that evaluates the selector function of the interaction $\alpha$ at the states of the nodes interacting at time $T-1$ through the link $ij$,
\begin{equation}
    F_n(X_{T-1}, X) = f_{\alpha}(s_i,s_j,s_i',s_j') \, .
    \label{eq:interaction_constraint_functions}
\end{equation}
This essentially indicates whether nodes interacting through the link $ij$ have carried out an interaction $\alpha$. Similarly, the identity selector functions define a single constraint function $F_n(X_{T-1}, X)$ that counts how many of the non-interacting nodes remain in the same state,
\begin{equation}
    F_n(X_{T-1}, X) = \sum_{l \neq i,j} \delta(s_l, s_l') \, .
    \label{eq:identity_constraint_functions}
\end{equation}

Having established the constraint functions of the process, we can write
\begin{equation}
    \begin{aligned}
        \sum_n \lambda_n F_n(X_{T-1}, X) =& \\
        \sum_{\alpha} \lambda_{\alpha} f_{\alpha} (s_i,s_j&,s_i',s_j') + \sum_{l \neq i,j} \lambda_I \delta(s_l,s_l') \, .
    \end{aligned}
    \label{eq:linear_combination_constraints}
\end{equation}
As the chosen constraint functions depend only on the last three times of the trajectory, the transitions they produce depend only on these variables, that is $\rho(X|X_{T-1}) = \rho(\vec{s}' | ij, \vec{s})$. Because the linear combination of constraints that define the transitions, namely \cref{eq:linear_combination_constraints}, presents a sum over independent nodes (with the exception of the two interacting through $ij$), the transition probabilities can be factorised into a product of independent node transition probabilities (again, with the exception of the two interacting ones),
\begin{equation}
    \rho(\vec{s}'|ij,\vec{s}) = B(s_i',s_j'|s_i,s_j) \prod_{l \neq i,j} C(s_l'|s_l)
    \label{eq:new_state_transitions}
\end{equation}
where
\begin{equation}
    \begin{aligned}
        B(s',r'|s,r) &= \frac{ e^{- \sum_\alpha \lambda_\alpha f_\alpha (s,r,s',r') }}{ Z_B(s,r)} \\
        Z_B(s,r) :&= \sum_{s',r'} e^{-\sum_\alpha \lambda_\alpha f_\alpha(s,r,s',r')}\\
        C(s'|s) &= \frac{ e^{ - \lambda_I \delta(s,s')} }{ Z_C(s) } \\
        Z_C(s) :&= \sum_{s'} e^{-\lambda_I \delta(s,s')} \,.
    \end{aligned}
    \label{eq:interactions_and_unchanged}
\end{equation}

As transition probabilities of non-interacting node states are given by $C$ and we have assumed that non-interacting nodes do not change states, we would like these transition probabilities to be $C(s|s') = \delta(s',s)$, that is $1$ if the states are the same and $0$ otherwise. Note that if there are $N_S$ states, $C(s|s) = (1 + (N_S - 1)e^{\lambda_I})^{-1}$ and $C(s'\neq s|s) = (N_S - 1 + e^{-\lambda_I})^{-1}$, so this can be achieved with $\lambda_I \rightarrow -\infty$. 

Just as $C$ establishes the transition probabilities of non-interacting nodes, $B$ establishes those of the interacting pair. To see that $B$ as defined through the chosen constraints can obtain any transition probability for the different interactions (as in the simulations), note that $B$ can be viewed as associating the value $e^{- \sum_\alpha \lambda_\alpha f_\alpha (s,r,s',r')}$ to each column $s,r$ and row $s',r'$ in the interaction matrix, and then normalising each column $s,r$ by the sum of its values $Z_B(s,r)$. Moreover, the associated values to transitions where change takes place depends only on the interaction $\alpha'$ associated it (instead of the transition itself) as only the corresponding selector function takes a value of $1$ while all the others $0$, $e^{- \sum_\alpha \lambda_\alpha f_\alpha (s,r,s',r')} = e^{-\lambda_{\alpha'}}$. For transitions where no change takes place, $(s,r) = (s',r')$, all selector functions take the value $0$, so the associated value to no change interactions is always $e^{- \sum_\alpha \lambda_\alpha f_\alpha (s,r,s,r)} = 1$. This way, non-negative values $v_\alpha = e^{-\lambda_\alpha}$ are associated to each interaction just as described before, with $v_{\text{n}} = 1$ for transitions where no change takes place, and allowing for $v_F \rightarrow 0$ in the appropriate limit of $\lambda_F$.

With the expression for transitions given an interacting pair of nodes of \cref{eq:new_state_transitions} and the definitions of \cref{eq:interactions_and_unchanged}, the joint probability of two consecutive population states is given by
\begin{equation}
    \begin{aligned}
        &\rho(\vec{s}, \vec{s}') = \sum_{ij} \rho(\vec{s},ij, \vec{s}') = \sum_{ij} \rho(\vec{s}'|ij, \vec{s}) \rho(ij | \vec{s}) \rho(\vec{s}) \\
    \end{aligned}
    \label{eq:population_successive_probabilities}
\end{equation}
The conditional probability $\rho(ij | \vec{s})$ reflects how a link is chosen given the state of the population. In general, one might include a dependence on the states of each node or an explicit time dependence, but as discussed for the cases presented here, we assume that this selection is uniform throughout the links (and therefore independent of the states) and constant in time, that is $\rho(ij | \vec{s}) = \rho(ij) = P_G(i,j)$. Combined with the expression for $\rho(\vec{s}' | ij, \vec{s})$ from \cref{eq:new_state_transitions}, the probability of two successive population states becomes
%\begin{widetext}
\begin{equation}
    \begin{aligned}
        \rho(\vec{s}', \vec{s}) 
        &= \sum_{i,j} P_G(i,j) \left[ B(s_i', s_j'|s_i,s_j) \prod_{l \neq i,j} \delta(s_l,s_l') \right] \rho(\vec{s}) \, .
    \end{aligned}
    \label{eq:population_transition_probabilities}
\end{equation}
%\end{widetext}
where we can sum over all pairs of nodes $i,j$ instead of over links $ij$ in the last expression because terms corresponding to disconnected pairs of nodes are weighted by $P_G(i,j) = 0$.

Let us now focus our attention on the probabilities of states of individual nodes instead of the entire population. If we have the distribution of states at a particular time $\rho(\vec{s})$, the probability $\rho_k(s)$ of a node $k$ being in a particular state $s$ can be calculated by summing the probabilities of all population states where $s_k = s$. This can be expressed as
\begin{equation}
    \rho_k(s) = \sum_{\vec{s}} \delta(s_k, s) \rho(\vec{s}) \, ,
\end{equation}
essentially the same average value of the delta function in \cref{eq:state_probability_estimation} but measured over the distribution instead of over realisations. 

Likewise, from the joint distribution of two consecutive population states $\rho(\vec{s},\vec{s}')$, we can sum over probabilities of all states where $s_k = s$ and $s_k'=s'$, namely
\begin{equation}
    \begin{aligned}
        \rho_k(s,s') &= \sum_{\vec{s}, \vec{s}'} \delta(s_k, s) \delta(s_k',s') \rho(\vec{s}, \vec{s}') \, . \\
    \end{aligned}
    \label{eq:consecutive_states}
\end{equation}
Replacing the expression for the probability of two consecutive states from \cref{eq:population_transition_probabilities}, we obtain
\begin{widetext}
\begin{equation}
    \begin{aligned}
        \rho_k(s,s') = \sum_{i,j} P_G(i,j) \sum_{\vec{s}} \delta(s_k,s) \left\{ \sum_{\vec{s'}} \delta(s_k',s') \left[ B(s_i',s_j'|s_i,s_j) \prod_{l \neq i,j} \delta(s_l,s_l') \right] \right\} \rho(\vec{s}) \, .
    \end{aligned}
    \label{eq:consecutive_states_general}
\end{equation}
\end{widetext}
To simplify notation, it is useful to define
\begin{equation}
    \begin{aligned}
         V_{ijk}(\vec{s},s') :&= \sum_{\vec{s'}} \delta(s'_k,s') \left[ B(s_i',s_j'|s_i,s_j) \prod_{l \neq i,j} \delta(s_l',s_l) \right] \\
         W_{ijk}(s,s') :&= \sum_{\vec{s}} \delta(s_k,s) V_{ijk}(\vec{s},s') \rho(\vec{s}) \, .
    \end{aligned}
    \label{eq:contributions_to_transitions}
\end{equation}

This way, \cref{eq:consecutive_states_general} tells us that the probability of two successive states at a node $k$ can be interpreted as given by contributions $W_{ijk}$ from each possible pair of nodes $i,j$ in the network weighted by their probability of interacting $P_G(i,j)$,
\begin{equation}
    \rho_k(s,s') = \sum_{i,j} P_G(i,j) W_{ijk}(s,s')  \, .
    \label{eq:node_transitions_by_node_pairs}
\end{equation}
Each of these contributions $W_{ijk}$, in turn, is an average over contributions $V_{ijk}(\vec{s},s')$ from each possible population state $\vec{s}$ before the interaction, weighted by the probability $\delta(s_k,s) \rho(\vec{s})$ of that population state, but only contributing if the desired initial state $s = s_k$ of the node $k$ is found. We will now proceed to simplify the expressions for these contributions.

Note that because we assume no self-loops, only pairs of different nodes $i \neq j$ can have $P_G(i,j) \neq 0$ and therefore contribute to the probability of consecutive states at $k$. The pair of interacting nodes $i,j$ must then be either both different from $k$, $i \neq k \neq j$, have $i = k \neq j$, or $i \neq k = j$. In each of these cases, the sum over population states $\vec{s'}$ after the interaction in the contribution $V_{ijk}$ from an initial state $\vec{s}$ can be expressed as a sum over possible states $s_m$ of each node $m$. However, the terms in the expression of $V_{ijk}$ must be arranged differently depending on the specific case.

For a pair of nodes $i,j$ both different from $k$, then
\begin{equation}
    \begin{aligned}
        V_{ijk}&(\vec{s},s') =\sum_{m} \sum_{s_m'} \delta(s_k',s') \left[ B(s_i',s_j'|s_i,s_j) \prod_{l \neq i,j} \delta(s_l,s_l') \right] \\
        =& \sum_{m} \sum_{s_m'} \delta(s_k',s') \delta(s_k,s_k') \left[ B(s_i',s_j'|s_i,s_j) \prod_{l \neq i,j,k} \delta(s_l,s_l') \right] \\
          =&\left( \sum_{s_k'} \delta(s_k',s')  \delta(s_k, s_k')\right) \left( \sum_{s_i',s_j'} B(s_i',s_j'|s_i,s_j) \right) \\
          & \times \prod_{l \neq i,j,k} \left( \sum_{s_l'} \delta(s_l,s_l') \right) \\
          =& \delta(s_k, s')
    \end{aligned}
\end{equation}
where in the first expression we sum over final states of each node rather than population states. In the second, the identity transition for the node $k$, $\delta(s_k',s_k)$, from the non-interacting nodes $l \neq i,j$ has been grouped with $\delta(s_k',s')$ as they depend on the states of the same node. We then factorise the sum over components of the final state into independent summations over components of the node $k$, the interacting nodes $i$ and $j$ together, and all other nodes $l \neq i,j,k$. Finally, we note that, except for the term corresponding to $k$, these independent sums are over final states of transition probabilities, and therefore are equal to $1$. In the term corresponding to $k$, the $\delta(s_k',s')$ evaluates $\delta(s_k,s_k')$ at $s_k' = s'$, yielding the final result. 

With this, the contribution $W_{ijk}(s,s')$ of any pair of nodes $i,j$ that are both different from $k$ is
\begin{equation}
    W_{ijk}(s,s') = \sum_{\vec{s}} \delta(s_k,s) \delta(s_k, s') \rho(\vec{s}) = \delta(s,s')\rho_k(s) \, .
    \label{eq:W_ijk}
\end{equation}

If, on the other hand, the node $k$ matches $i$, the contribution from each initial state weighted by the probability of the state and the Kronecker delta, is
\begin{equation}
    \begin{aligned}
        V_{kjk}&(\vec{s},s')= \sum_{m} \sum_{s_m'} \delta(s_k',s') B(s_k',s_j'|s_k,s_j) \prod_{l \neq k,j} \delta(s_l,s_l') \\
        =& \left( \sum_{s_k', s_j'} \delta(s_k', s')B(s_k',s_j'|s_k,s_j)  \right) \prod_{l \neq k,j} \left( \sum_{s_l'} \delta(s_l,s_l') \right) \\
        =& \sum_{s_j'} B(s',s_j'|s_k,s_j) = \sum_{r'} B(s',r'|s_k,s_j) \, .
    \end{aligned}
    \label{eq:ki_contribution}
\end{equation}
This time, the term corresponding to the interacting nodes $i = k$ and $j$ are grouped with the $\delta(s_k',s')$. Terms corresponding to all other nodes $l \neq k,j$ are then normalised, while the term corresponding to the interacting nodes is evaluated at $s_k' = s'$. Note that because transitions of interacting nodes depend only on the states, the sum over final states of $j$ can be expressed as $\sum_{s_j'} B(s',s_j'|s_k,s_j) =\sum_{r'} B(s',r'|s_k,s_j)$. 

The contribution $W_{ijk}(s,s')$ when $k = i$ is at the source of the link connecting it to another node $j \neq k$ it interacts with is then
\begin{equation}
    \begin{aligned}
        W_{kjk}&(s,s') = \sum_{\vec{s}} \delta(s_k, s) \sum_{r'} B(s', r' | s_k, s_j) \rho(\vec{s}) \\
        =&
        \sum_{\vec{s}} \delta(s_k, s) \sum_{r', r} \delta(s_j, r) B(s', r' | s_k, s_j) \rho(\vec{s}) \\
        =& \sum_{r', r} B(s', r' | s, r) \rho_{kj}(s, r) \, ,
    \end{aligned}
    \label{eq:W_kjk}
\end{equation}
where we recognise the joint probability $\rho_{kj}(s,r) = \sum_{\vec{s}} \delta(s_k,s)\delta(s_j,r) \rho(\vec{s})$ of the pair of nodes $k,j$ in states $s,r$ respectively, analogous to \cref{eq:state_pairs_probability_estimation}.
Similarly, when the node $k$ matches the target of the link defining the interaction $k = j$
\begin{equation}
    W_{ikk}(s,s') = \sum_{r', r} B(r', s' | r, s) \rho_{ki}(s, r)
    \label{eq:W_ikk}
\end{equation}

Thus, if we express \cref{eq:node_transitions_by_node_pairs} as sums over pairs of nodes that are both different, or one different from $k$, and use \cref{eq:W_ijk,eq:W_kjk,eq:W_ikk} to introduce the contributions in each case, the probability of two consecutive node states becomes
\begin{widetext}
\begin{equation}
    \begin{aligned}
        \rho_k (s,s')
        = \left( 1 - \sum_i P_G(i,k) + P_G(k,i) \right) \delta(s,s') \rho_k(s) + \sum_i \sum_{r', r} \Big{[} P_G(i,k) B(r',s'|r,s) + P_G(k,i)  B(s',r'|s,r) \Big{]} \rho_{ki}(s, r) \, .
    \end{aligned}
    \label{eq:single_cell_dynamics}
\end{equation}
\end{widetext}
The first term is the contributions from pairs $i,j \neq k$ of interacting nodes both different from $k$. Because all these contribute an amount that does not depend on the pair of interacting nodes (\cref{eq:W_ijk} does not depend on $i$ or $j$), then the sum over probabilities $\sum_{i,j \neq k} P_G(i,j)$ of such links being selected can be factored out. This sum is then related to the probabilities of node $k$ interacting because the probability of picking a pair of nodes is normalised $\sum_{i \neq k} P_G(i,k) + \sum_{j \neq k} P_G(k,j) + \sum_{i,j \neq k} P_G(i,j) = 1$. The second term considers all contributions from nodes $i$ interacting with $k$ and vice versa according to \cref{eq:W_kjk,eq:W_ikk}.

Thus the joint probability of a node in two successive states depends on the joint probability of two nodes in two states before the interaction. While the probability of the state of each node can be updated from the joint probability of successive states by marginalising over the first, $\rho_k(s') = \sum_s \rho_k(s,s')$, the joint probability of pairs of node states cannot. This means that for the joint probability of successive states in the next step, the new probability of the state of each node will not be enough to calculate the new joint probability of two successive states and perform another update of the probability of single-node states. If we attempt to carry out the same calculation to obtain how to update the joint probabilities of pairs of states, the result depends on the joint probability of three node states. This introduces a hierarchical dependence that eventually requires considering all node states of the network simultaneously. 

The simplest way to avoid considering all node states of the network simultaneously is to introduce what we refer to as the independent node approximation, where the dynamics depends on the probabilities of states in two independent nodes instead of their joint probabilities. This consists of replacing $\rho_{ki}(s,r)$ with $\rho_k(s) \rho_i(r)$ in \cref{eq:single_cell_dynamics}, resulting in the joint probability of pairs of node states not being needed to calculate the probability of pairs of successive states.  Note that while one might indeed choose initial conditions where this is the case, i.e. nodes are independent $\rho_{ki}(s,r) = \rho_k(s) \rho_i(r)$, the pairwise transitions $B$ will introduce dependence among nodes already at the second step, meaning that $\rho_{ki}(s,r) \neq \rho_k(s)\rho_i(r)$. What we propose is a different dynamical equation approximating the original. In this approximation, once these probabilities of successive states are used to update the probability of the state of each independent node, they are again enough to obtain the probability of pairs of successive states and repeat the update as the dynamics no longer depend on the joint distribution.

\subsection{Comparing simulations and maximum caliber}

Maximum caliber gives a precise formulation on how to update the probabilities of individual node states. Unfortunately, it tells us that we need more than just the probabilities of individual nodes to carry out this update. If this equation captures the evolution of the probabilities of node states, then it will be worthwhile to develop methods to solve it exactly in future research. However, our goal in this paper is to verify whether \cref{eq:single_cell_dynamics} actually holds for the dynamics studied. We can do this without solving the equation simply by verifying that when we evaluate the right and left hand sides using the probabilities measured from simulations, these are in fact the same.

Additionally, we directly compare the probabilities of individual node states measured over time with the result of using \cref{eq:single_cell_dynamics} in the independent node approximation. In this case, we evaluate the updates of an initial distribution $\rho_o(\vec{s})$ that takes the value $\rho_o(\vec{s}) = 1$ if $\vec{s} = \vec{s}(0)$ is the initial condition used for simulations, and $0$ otherwise. This is intended to provide some insight on the range of validity of an assumption that can be used to make estimations.

\subsection{Relation to population dynamics models}
\label{sec:methods:pop_dyn}

The independent node approximation is also useful to relate the proposed dynamics of the different processes considered to existing population models that do not account for an underlying network. These models define how the total number of individuals of different groups evolve over time, assuming that any individual in the population can interact with any other. This hypothesis is known as complete mixing, introduced here through a fully connected network. It allows us to derive the logistic population model from competition-limited population growth, the Lotka-Volterra model from the predator prey system, and the SIR model for disease spreading. 

For this, we consider how \cref{eq:single_cell_dynamics} is further simplified from the independent node approximation $\rho_{ki}(s,r) = \rho_k(s)\rho_i(r)$ when links are chosen uniformly from a fully connected network $P_G(i,j) = (1 - \delta(i,j))/(N(N-1))$. We apply this to obtain the evolution of $X_{\sigma} = \sum_i \rho_i(\sigma)$, the expected population of nodes in state $\sigma$ in the network. For the evolution of the population, we can write the difference between the population at two successive time steps as
\begin{equation}
    \Delta X_{\sigma} = \sum_{i,s} \rho_i (\sigma, s) - X_{\sigma} \, .
\end{equation}

In the independent node approximation, for a fully connected network and reflection-symmetric interactions $B(s',r'|s,r) = B(r',s'|r,s)$ (note that this is the case for the interaction tables used in our examples), we can use \cref{eq:single_cell_dynamics} to write
\begin{equation}
    \begin{aligned}
        \sum_{i,s} \rho_i(\sigma, s) =& \sum_{i,s} \left\{ \left(1 - \sum_j \frac{2(1 - \delta_{ij})}{N (N - 1)} \right) \delta(\sigma, s) \rho_i(s) \right. \\
         + \sum_{j,r',r} & \left. \frac{2 B(\sigma, r' | s, r) \left( 1 - \delta_{ij} \right) }{N (N - 1)} \rho_i(s) \rho_j(r) \right\} \\
        =& \left(1 - \frac{2}{N} \right) X_{\sigma} \\
        + \sum_{r',s,r} &\frac{2 B(\sigma, r' | s, r) }{N (N - 1)} \left( X_s X_r - \sum_i \rho_i(s) \rho_i(r) \right) \\ 
    \end{aligned}
\end{equation}
from where, defining $C(\sigma | s,r) := \sum_{r'} B(\sigma, r' | s,r)$ 
\begin{equation}
    \Delta X_{\sigma} = - \frac{2}{N} X_{\sigma} + \sum_{s,r} \frac{2 C(\sigma | s, r) }{N (N - 1)} \left( X_s X_r - \sum_i \rho_i(s) \rho_i(r) \right) \, .
    \label{eq:population_dynamics_with_empty}
\end{equation}

Because the state $s = 0$ always represents an empty node in the cases we consider, we seek to express the evolution of a population $\sigma$ in terms of non-empty node states only. This can be done because the probabilities of states in each node are normalised $\sum_s \rho_i(s) = 1 ~ \forall ~ i$ and because the total number of nodes is fixed, $N = \sum_{i,s} \rho_{i}(s) = \sum_s X_s$. Assuming Greek letters imply non-empty states, empty states can be replaced by $\rho_i(0) = 1 - \sum_{\alpha} \rho_i(\alpha)$ and $X_0 = N - \sum_{\alpha} X_{\alpha}$. Expanding the sum over pairs of initial node states in \cref{eq:population_dynamics_with_empty} into terms involving empty and non-empty states

\begin{equation}
    \begin{aligned}
        \Delta X_{\sigma} &= - \frac{2}{N} X_{\sigma} + \sum_{\alpha,\beta} \frac{2 C(\sigma | \alpha, \beta) }{N (N - 1)}  \left( X_{\alpha} X_{\beta} - \sum_i \rho_i(\alpha) \rho_i(\beta) \right) \\
         +&  \sum_{\alpha} 2 \frac{C(\sigma | 0, \alpha) + C(\sigma | \alpha, 0) }{N (N - 1)} \left( X_0 X_\alpha - \sum_i \rho_i(0) \rho_i(\alpha) \right) \\
         +&  \frac{2 C(\sigma | 0, 0) }{N (N - 1)} \left( {X_0}^2 - \sum_i {\rho_i(0)}^2 \right) \, . \\
    \end{aligned}
\end{equation}
Replacing the empty states in terms of non-empty ones, and defining the constants
\begin{equation}
    \begin{aligned}
        \gamma_\sigma &= C(\sigma|0,0) \\
        \gamma_\sigma(\alpha) &= C(\sigma | 0, \alpha) + C(\sigma | \alpha, 0) - 2 C(\sigma|0,0) \\
        \gamma_\sigma(\alpha,\beta) &= C(\sigma | \alpha, \beta) - C(\sigma | \alpha, 0) - C(\sigma | 0, \beta) + C(\sigma | 0, 0)
    \end{aligned}
\end{equation}
that depend only on the values of the transition matrix $B$, we obtain
\begin{equation}
    \begin{aligned}
        \frac{\Delta X_{\sigma}}{2} &= \gamma_\sigma - \frac{X_{\sigma}}{N} + \sum_{\alpha} \gamma_{\sigma}(\alpha)\frac{X_\alpha}{N}  \\
        +&\sum_{\alpha,\beta} \gamma_{\sigma}(\alpha,\beta)
        %\Bigg\{ \bigg[ C(\sigma | \alpha, \beta) - C(\sigma | \alpha, 0) - C(\sigma | 0, \beta) + %C(\sigma | 0, 0) \bigg] \\
        %& ~~~~~~~~ \times 
        \frac{ X_{\alpha} X_{\beta} - \sum_i \rho_i(\alpha) \rho_i(\beta) }{N (N - 1)} \, .%\Bigg\} \, .
        %\frac{\Delta X_{\sigma}}{2} &= C(\sigma | 0, 0) - \frac{X_{\sigma}}{N} \\
        %+& \sum_{\alpha} \bigg[ C(\sigma | 0, \alpha) + C(\sigma | \alpha, 0) - 2 C(\sigma | 0, 0) \bigg] \frac{X_\alpha}{N}  \\
        %+&\sum_{\alpha,\beta} \Bigg\{ \bigg[ C(\sigma | \alpha, \beta) - C(\sigma | \alpha, 0) - C(\sigma | 0, \beta) + C(\sigma | 0, 0) \bigg] \\
        %& ~~~~~~~~ \times \frac{ X_{\alpha} X_{\beta} - \sum_i \rho_i(\alpha) \rho_i(\beta) }{N (N - 1)} \Bigg\} \, .
    \end{aligned}
    \label{eq:population_dynamics_methods}
\end{equation}    
Finally, we note that the term $\sum_i \rho_i(\alpha) \rho_i(\beta) / N (N - 1)$ will always be the smallest order in the dynamics as $\rho_i(\alpha) \leq 1 ~ \forall ~ \alpha, i$. Therefore we ignore this contribution for the cases considered, resulting in the expression
\begin{widetext}
\begin{equation}
    \begin{aligned}
        \frac{\Delta X_{\sigma} }{2} =& \gamma_\sigma - \frac{X_{\sigma}}{N} + \sum_{\alpha} \gamma_\sigma(\alpha) \frac{X_\alpha}{N} + \sum_{\alpha,\beta} \gamma_\sigma(\alpha,\beta) \frac{ X_{\alpha} X_{\beta} }{N (N - 1)} \, .
%        \frac{\Delta X_{\sigma} }{2} =& C(\sigma | 0, 0) - \frac{X_{\sigma}}{N} + \sum_{\alpha} \bigg[ C(\sigma | 0, \alpha) + C(\sigma | \alpha, 0) - 2 C(\sigma | 0, 0) \bigg] \frac{X_\alpha}{N}  \\
%        & +\sum_{\alpha,\beta} \bigg[ C(\sigma | \alpha, \beta) - C(\sigma | \alpha, 0) - C(\sigma | 0, \beta) + C(\sigma | 0, 0) \bigg] \frac{ X_{\alpha} X_{\beta} }{N (N - 1)} \, .
    \end{aligned}
    \label{eq:population_dynamics}
\end{equation}    
\end{widetext}

\section{Results}

\subsection{Competition-limited population growth}

\begin{figure}[h!]
    \centering
    \includegraphics[width=\linewidth]{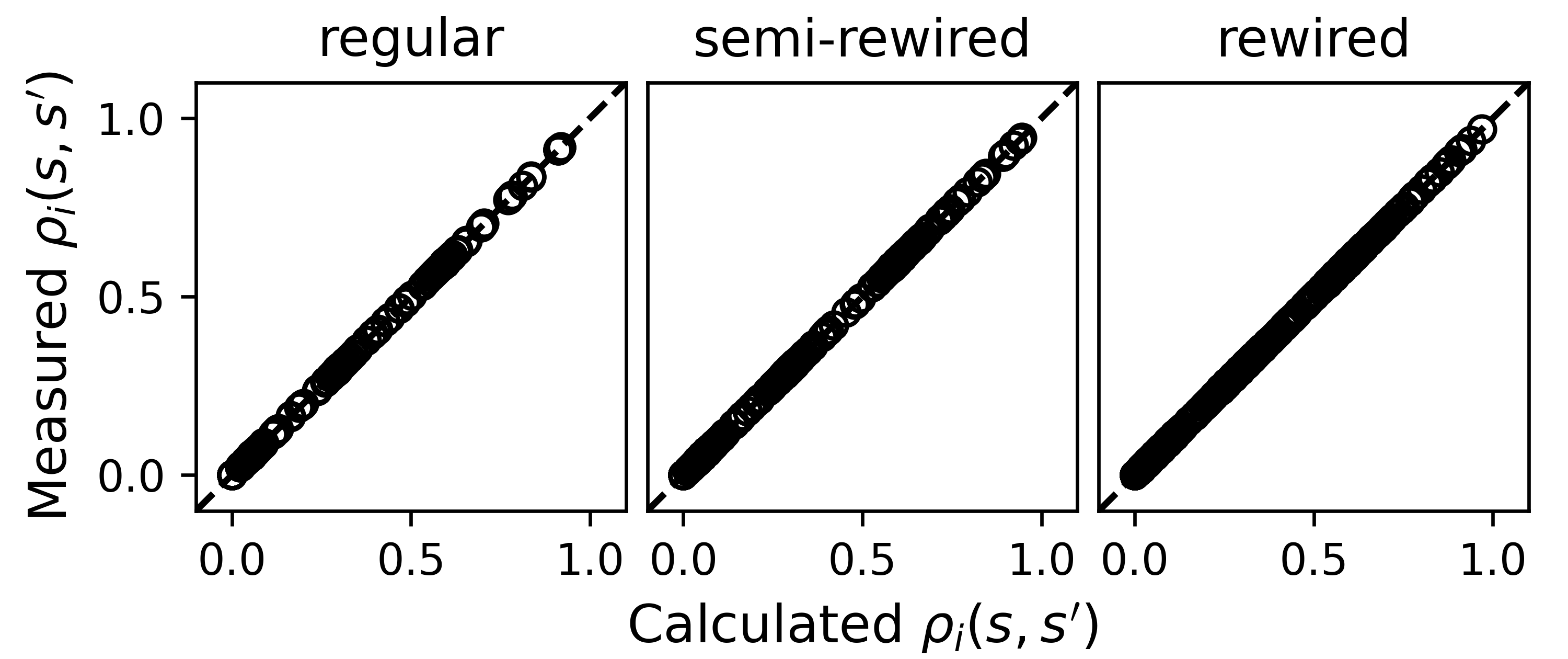}
    \caption{Probabilities of successive states measured from simulations as a function of values calculated according to maximum caliber for competition-limited population growth. Each subplot shows a particular network and each circular marker in a given subplot corresponds to a pair of consecutive states at a particular node in the corresponding network and time elapsed in the evolution of the competition-limited population growth process. As all points fall on the identity, the result of maximum caliber holds for all these conditions.}
    \label{fig:logistic_dynamics}
\end{figure}

In \cref{fig:logistic_dynamics} we show the joint probabilities pairs of successive states $\rho_i(s,s')$ measured directly from simulations as a function of the same probability calculated according to \cref{eq:single_cell_dynamics}. Each subplot corresponds to one of the networks tested, and each circular marker corresponds to a particular pair of successive states found at a given node at a certain point in time of the simulation. As all points fall on the identity (dashed line), the left (measured) and right (calculated) hand sides of \cref{eq:single_cell_dynamics} are equal and therefore the equation holds for all conditions tested in this process.

Since the dynamics of maximum caliber hold exactly throughout the evolution of the process on the networks considered, it is worthwhile to explore how accurately the independent node approximation captures the evolution of probability. While it is only an approximation, it allows to update an initial distribution an arbitrary amount of steps, while the exact dynamics allows only for a single step. 

\begin{figure}[h!]
    \centering
    \includegraphics[width=\linewidth]{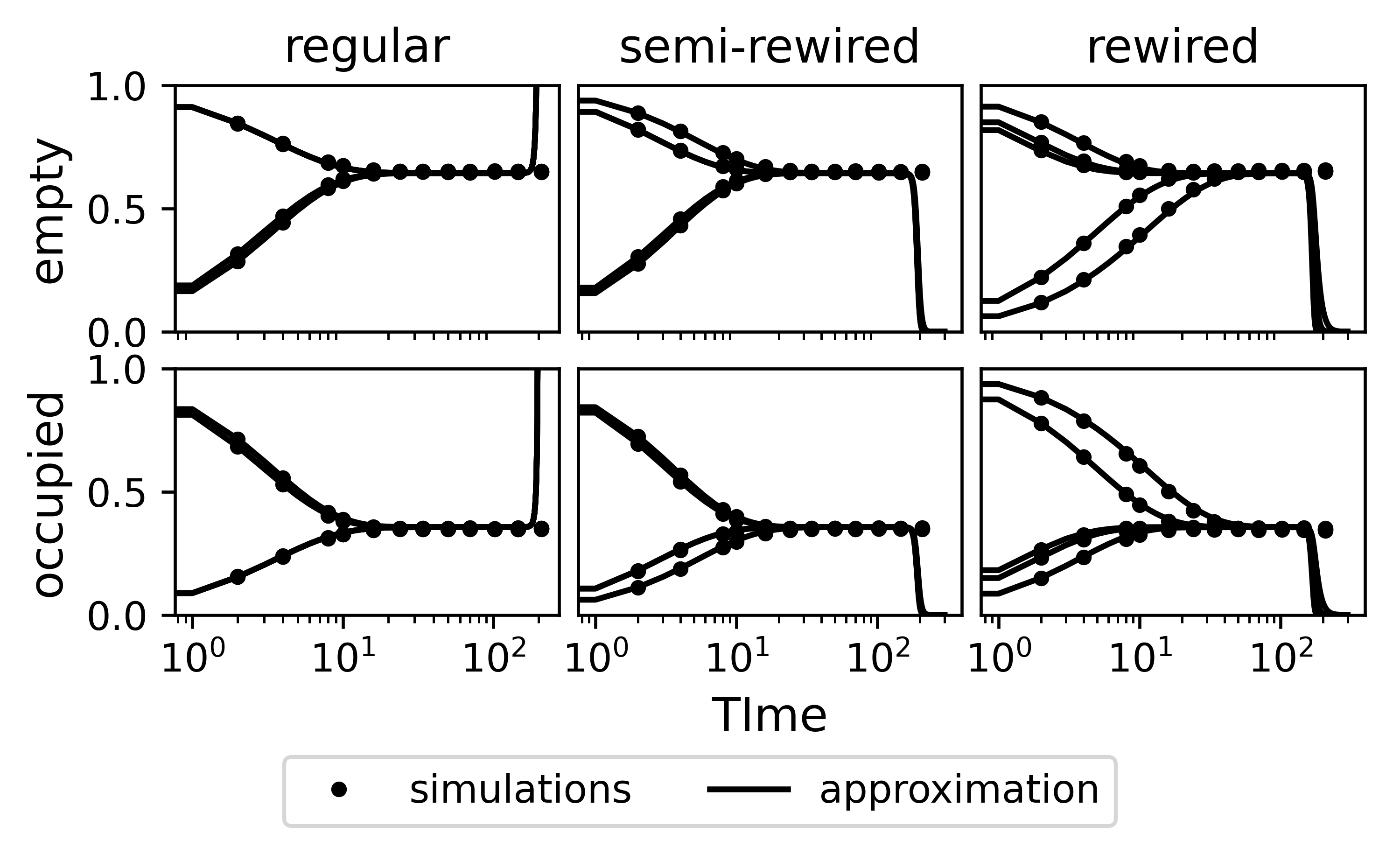}
    \caption{Probabilities of a particular state of competition-limited population growth (given by the subfigure row) in select nodes of the different networks (subfigure column) as a function of time. Circular markers are measured from simulations while full lines are estimated with maximum caliber. The approximation holds for an initial transitory period of the dynamics but then departs abruptly.}
    \label{fig:logistic_approximation}
\end{figure}

As a test of the range of validity of the independent node approximation, in \cref{fig:logistic_approximation}, we show the probabilities of different states in the networks used as a function of time. Each subfigure shows the probabilities of a particular state (according to the subfigure row) in a particular network (given by the subfigure column) for select nodes in the network. Circular markers show the probability of the state in each node of the network  according to simulations, while full lines show the value calculated according to maximum caliber in the independent node approximation. Note that the approximation holds precisely for an initial transient period but departs abruptly. 

The agreement between results from the approximation and those from simulations, at least at short times, motivates its use to obtain the evolution of the total expected population of individuals in the system $X_1 = \sum_i \rho_i(1)$. In particular, the logistic model of population growth describes a population that increases until competition for a limited carrying capacity of the system saturates it, assuming complete mixing \cite{tsoularis2002analysis}. We therefore expect the population dynamics for the competition-limited population growth on a fully connected network, i.e. \cref{eq:population_dynamics} with constants defined by the transitions in \cref{tab:logistic_interactions}, to yield a similar population dynamics.

The main remaining difference between the assumptions of the logistic model and competition-limited population growth is that for the former, an population of zero individuals is in equilibrium (albeit unstable), meaning it remains with zero individuals. In the latter, on the other hand, a non-zero probability for spontaneous births would allow a single individual to appear spontaneously. Assuming the probability of a spontaneous birth is $0$ (that is, making spontaneous births forbidden transitions) we find from \cref{eq:population_dynamics} that the dynamics of the expected number of occupied nodes $X_1$ follows a logistic equation that describes saturating population growth limited by a carrying capacity of the system\cite{tsoularis2002analysis}, 
\begin{equation}
    \Delta X_1 = 2 \frac{c + r - d}{N (N - 1)} X_1 \left( (N-1) \frac{r - d}{c + r - d} - X_1 \right) \, .
\end{equation}

\subsection{Predator prey dynamics}

\begin{figure}[h!]
    \centering
    \includegraphics[width=\linewidth]{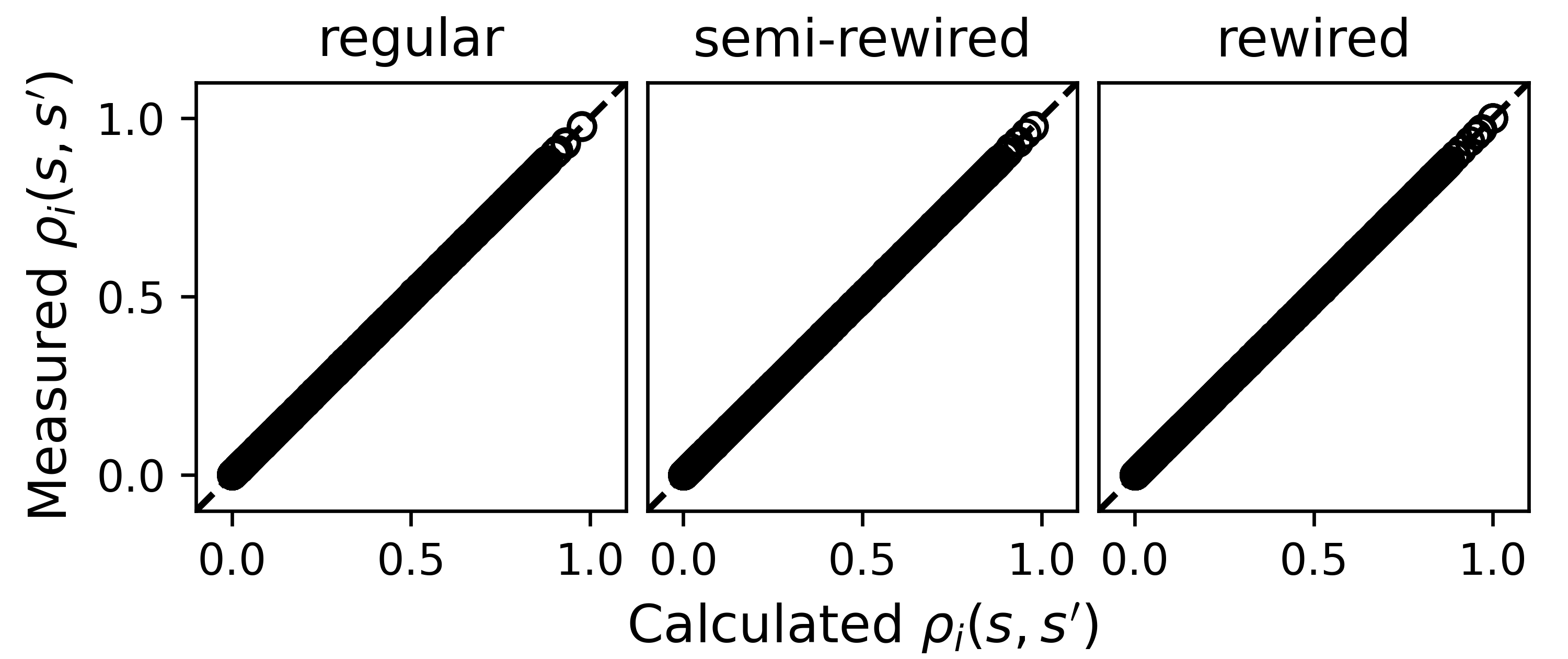}
    \caption{Probabilities of successive states measured from simulations as a function of values calculated from maximum caliber for the predator-prey system.  Each subplot shows a particular network and each circular marker in a given subplot corresponds to a pair of consecutive states at a particular node in the corresponding network and time elapsed in the evolution of the predator prey system. As all points fall on the identity, the result of maximum caliber holds for all these conditions.}
    \label{fig:rep_dynamics_lotka_volterra}
\end{figure}

As for the competition-limited population growth dynamics, we now show the probabilities of consecutive states measured directly from simulations as a function of their values as calculated according to \cref{eq:single_cell_dynamics}. In \cref{fig:rep_dynamics_lotka_volterra}, each subfigure corresponds to a particular network, and circular markers correspond to probabilities of a certain pair of successive states of the predator-prey system at a given node in the corresponding network and time of the evolution of the predator-prey system. As in the case of competition-limited population growth, the plotted points fall on the identity (dashed line), indicating that the dynamics obtained from maximum caliber is valid for all conditions tested in this process.

\begin{figure}[h!]
    \centering
    \includegraphics[width=\linewidth]{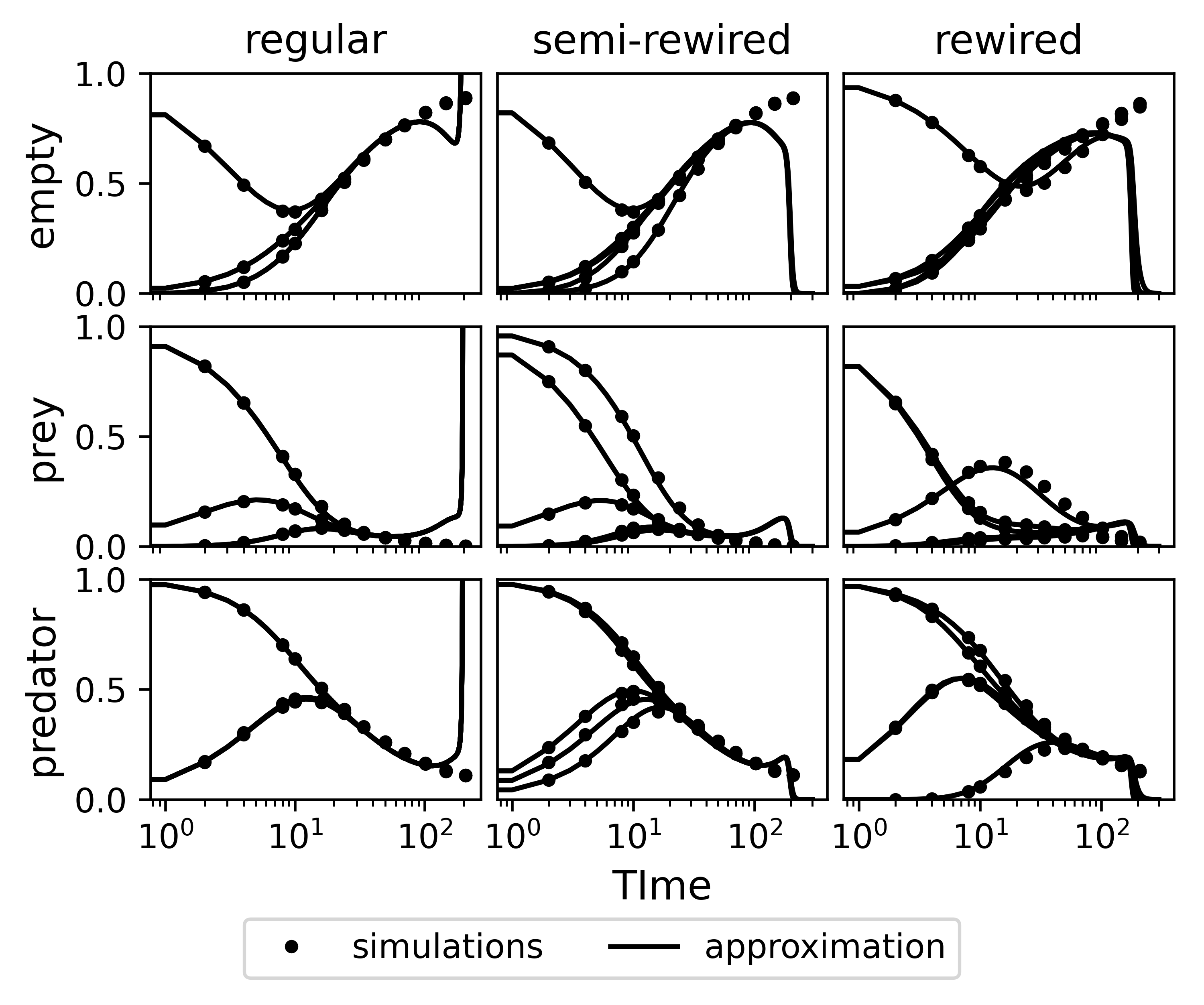}
    \caption{Probabilities of a particular state of predator-prey system (given by the subfigure row) in select nodes of the different networks (subfigure column) as a function of time. Circular markers are measured from simulations while full lines are estimated with maximum caliber. The approximation reproduces the simulation results for an initial transitory period but then departs, less abruptly than for competition-limited population growth.}
    \label{fig:lotka_volterra_apprixmation}
\end{figure}

We now proceed to introduce the independent node approximation. In \cref{fig:lotka_volterra_apprixmation} each subfigure corresponds to a state (given by the column in which the subfigure is located) and a network (given by the row). Each subfigure then shows the approximated evolution in full lines and the measurements from simulations in circular markers. Again, the approximation reproduces the simulated results for an initial transitory period. However, it is also followed by a departure of the approximation (less abrupt than in competition-limited population growth, but still sudden) from the results of simulations. 

Similarly to the logistic model, the Lotka-Volterra model captures the evolution of the population of prey and predators. It assumes complete mixing, which we capture by considering the population on a fully connected network, and does not allow for spontaneous births of prey and predators. Under these assumptions, i.e. $b_1 = b_2 = 0$, the population of prey $X_1$ and predators $X_2$ resulting from \cref{eq:population_dynamics} evolve according to the Lotka-Volterra model of predator-prey systems\cite{bunin2017ecological}
\begin{equation}
    \begin{aligned}
        \frac{\Delta  X_1} {2} =& (r_1 - d_1) \frac{X_1}{N} - (r_1 - d_1 + p + r_2) \frac{ X_1 X_2 }{N (N - 1)} \\
        &- (r_1 - d_1 + c_1) \frac{{X_1}^2}{N(N-1)} \\
        \frac{\Delta X_2}{2} =& -  d_2 \frac{X_2}{N} + (r_2 + d_2) \frac{X_1  X_2}{N (N - 1)} \\
        &+ (d_2 - c_2) \frac{{X_2}^2 }{N (N - 1)} \, .
    \end{aligned}
\end{equation}
While the most common Lotka-Volterra model does not include the terms proportional to ${X_1}^2$ and ${X_2}^2$, these are included in generalised Lotka-Volterra systems or can be eliminated by specific choices of the relation between parameters.

\subsection{Epidemic spreading}

\begin{figure}[h!]
    \centering
    \includegraphics[width=\linewidth]{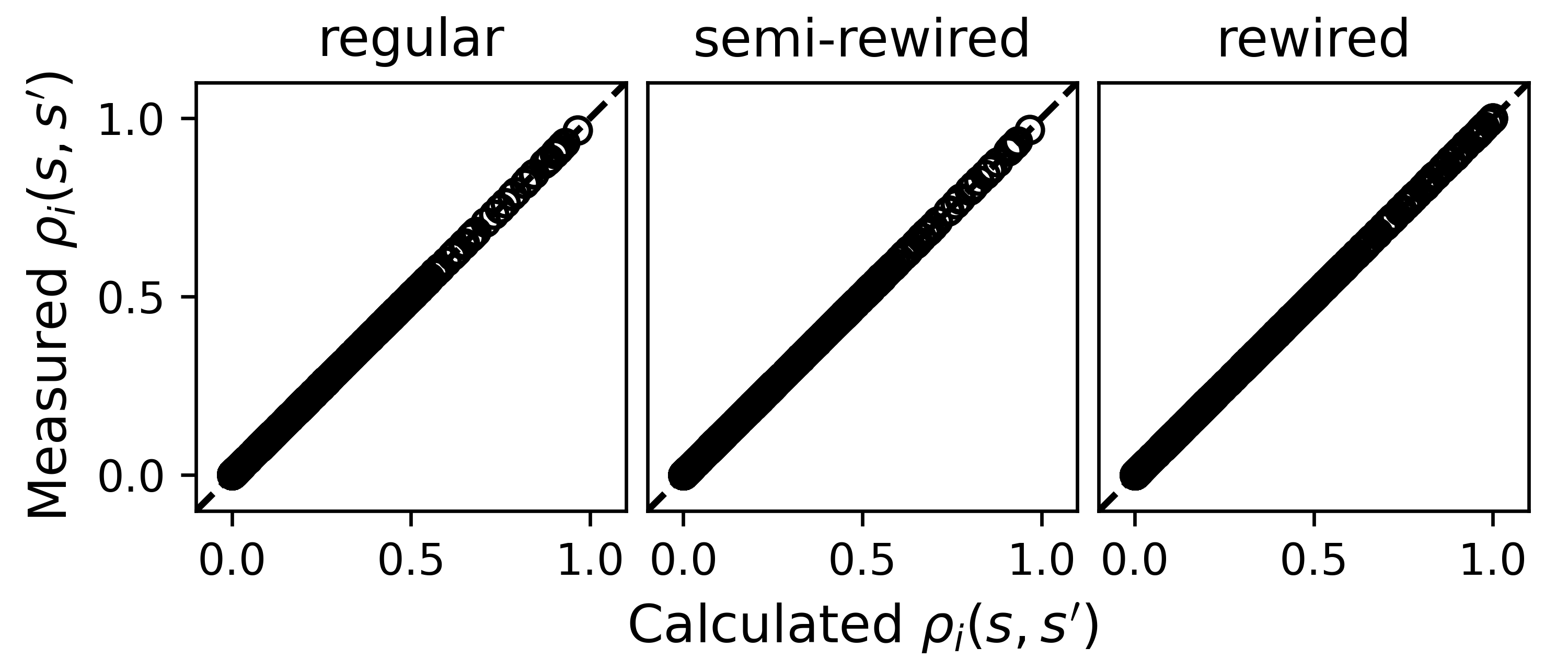}
    \caption{Measured probabilities of consecutive states as a function of values calculated according to maximum caliber for the epidemic spreading process. Each subfigure shows a particular network, and in it, markers are shown for each pair of states, nodes, and times tested. As points fall on the identity line, the dynamics resulting from maximum caliber holds for all of these conditions.}
    \label{fig:rep_dynamics_SIR}
\end{figure}

In \cref{fig:rep_dynamics_SIR} we show the probabilities of successive states as measured directly from simulations as a function of the probabilities calculated from maximum caliber according to \cref{eq:single_cell_dynamics}. Each subfigure shows a particular network, and probabilities in the subfigure are measured and calculated for all different nodes, transitions, and times in the epidemic spreading process. Once again, values fall on the identity line showing that maximum caliber correctly predicts the probabilities of transitions. 

\begin{figure}[h!]
    \centering
    \includegraphics[width=\linewidth]{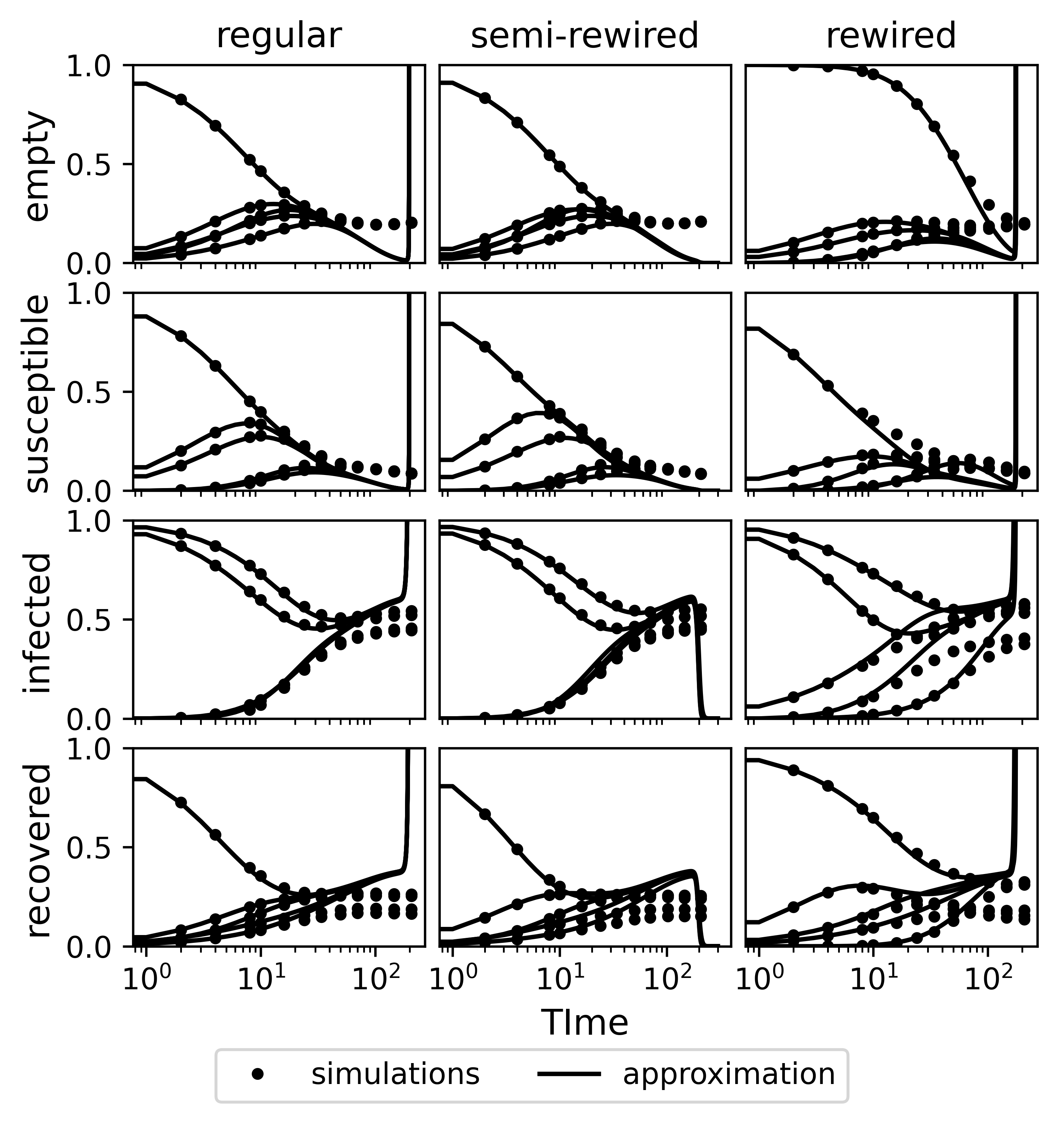}
    \caption{Probabilities of different states (row) in each node of the different networks used (column) as a function of time. Measurements from simulations are shown in circular markers and results from the approximation of maximum caliber in full lines. The approximation reproduces simulation results for an initial transitory period, departing from them earlier in the evolution of the system and also less abruptly.}
    \label{fig:SIR_apprixmation}
\end{figure}

In \cref{fig:SIR_apprixmation} each subfigure shows probabilities of each state (given by the row of the subfigure) in the different networks (given by the subfigure column) in select nodes of each network as a function of time. Probabilities calculated according to the independent node approximation are shown in full lines and the results from simulations in circular markers. The approximation holds for an initial transient period but departs from measurements before it did in the predator-prey model, but also less suddenly.

Now, assume that no birth or death dynamics ($r_i = d_i = 0$ and $r_{21} = r_{31} = 0$), recovered individuals do not become susceptible again $l = 0$, and the total population (expected amount of non-empty nodes) is fixed $\sum_{\alpha=1}^3 X_{\alpha} = N' \leq N$. These are assumptions of the SIR model of epidemic spreading\cite{newman2002spread} that, when introduced, simplify the dynamics of the population according to \cref{eq:population_dynamics}, that is on a fully connected network representing complete mixing, to that of the SIR model, namely
\begin{equation}
    \begin{aligned}
        \Delta X_1 / 2 &=  - c \frac{X_1 X_2}{ N (N - 1) } \\
        \Delta X_2 / 2 &=  c \frac{X_1 X_2}{ N (N - 1) } - h \left( 1 - \frac{N'}{N - 1} \right) \frac{X_2}{N} \\
        \Delta X_3 / 2 &= h \left( 1 - \frac{N'}{N - 1} \right) \frac{X_2}{N} \, .
    \end{aligned}
\end{equation} 

\section{Discussion and conclusions}

In this work, we have used the principle of maximum caliber to derive the dynamics of diffusion processes on networks. We have applied these dynamics to three specific processes, each on three different (albeit static) network structures, and compared them to the results from stochastic simulations. While we have not been able to integrate the exact resulting dynamics analytically or numerically, we have shown that the transitions predicted by the method match those measured from simulations. We have also introduced an approximation of this dynamics which can be integrated, finding that it matches results from simulations accurately in an initial phase of the evolution, but also departs abruptly. Finally, we have shown analytically that for the case of a fully connected network, the dynamics reduce to well-known diffusion models on homogeneous spaces such as logistic (competition limited) population growth, Lotka-Volterra (predator-prey) dynamics, and SIR (epidemic) spreading.   

The lack of a way to integrate the dynamics of the system emerges from a hierarchical dependence of the evolution of groups of nodes: the evolution of the probability of the state of a single node depends on joint probabilities of pairs of node states. The evolution of the joint probability of pairs of node states depends on the joint probabilities of triads of node states, and so on. We believe that this hierarchy is the reason behind the difficulty in treating dynamical processes on networks analytically in general and not just a fault of the method introduced here. However, the insight gained will hopefully allow to better focus future efforts in addressing the problem, for example developing better approximations and understanding their range of validity.

In other avenues of future research, the method used here can be extended in three main ways. First, here we have focused on population dynamics where each node is occupied by at most one individual, while many cases of interest involve more than one individual occupying a node. Second, we have considered discrete states and time, while continuous states and time are fundamental to many dynamical processes. Third, and perhaps most interesting, we have considered that the structure of the network is fixed, while allowing the structure to be updated depending on the states of nodes (just as the node states are updated depending on the structure in this case) could produce a framework to understand the interplay of the structure of a network and ongoing dynamics.

\section*{Acknowledgements}

We would like to thank Professor Guillermo Abramson for laying the foundations of the ideas presented in this        work.

\appendix

\section{Selector functions and moments of the distribution}
\label{app:moments}

In the case of a discrete-state process on an $N$-node network, binary states $\mathcal{S} = \{0,1\}$ are appealing for their analytical simplicity. For example, the number of nodes occupied by states $s_i = 1$ can simply be expressed as $X_1 = \sum_i s_i$, and is therefore related to the first moment of the distribution of states. The amount of nodes occupied by the state $0$ must then be $X_0 = N - X_1$. 

On the other hand, here we have used the Kronecker delta $\delta(s_i, \sigma)$ to count the number of nodes in a particular state $\sigma$, that is $N_\sigma = \sum_i \delta(s_i, \sigma)$. Note that, on the set of states $\mathcal{S} = \{0,1\}$, we can write $\delta(s_i, 1) = s_i$ (as the expressions on either side take the same values), and also $\delta(s_i, 0) = 1 - \delta(s_i, 1) = 1 - s_i$.

Moreover, for the states $\mathcal{S} = \{0,1,2\}$, we can express the Kronecker delta $\delta(s_i, \sigma)$ for $\sigma= 1$ as $\delta(s_i, 1)= -s_i(s_i - 2)$. This is a parabola with roots at $0$ and $2$, taking the value $1$ at $s_i = 1$. For $\sigma = 2$, we want a parabola with roots at $0$ and $1$ taking the value $1$ at $s_i = 2$, for which we can write $\delta(s_i,2) = s_i(s_i - 1)/2$. We can then write $\delta(s_i,0) = 1 - \delta(s_i,1) - \delta(s_i,2) = (s_i-1)(s_i - 2)/2$

If we then continue onto $S$ states $\mathcal{S} = [0,S) \subset \mathbf{Z}$, the Kronecker delta for each $\sigma > 0$ is a polynomial of order $S-1$ with roots at every integer in $\mathcal{S}$ except $\sigma$,
\begin{equation}
    \delta(s_i, \sigma) = K_\sigma \prod_{\substack{n \in \mathcal{S} \\ n \neq \sigma}} (s_i - n) \, ,
\end{equation}
where $K_\sigma = [(-1)^{S - \sigma - 1} \sigma! (S -\sigma - 1)!]^{-1}$ is a proportionality constant that ensures that the result is $1$ at $s_i = \sigma$. The point of this is that constraints involving the Kronecker delta of a single state can then be interpreted as constraints on linear combinations of the moments of the distribution of that state. Because the result of Lagrangian maximisation is invariant under linear combinations of imposed constraints, this is equivalent to constraining the moments directly. However, constraints on the moments of arbitrarily assigned indices are far less intuitive to interpret than the Kronecker delta.

Furthermore, the selector functions defined to detect each of the interactions in the process can be expressed as sums and products of Kronecker deltas. For example, the spontaneous birth and movement selector functions $f_b$ and $f_m$ in the competition-limited population growth processes (defined in \cref{eq:birth_selector,eq:movement_selector}) can be expressed as
\begin{equation}
    \begin{aligned}
        f_b(s',r',s,r) = \delta(s,0) \delta(r,0) [ & \delta(s',1)\delta(r',0) \\
        + &\delta(s',0)\delta(r',1) ] \\
        f_m(s',r',s,r) = \delta(s,0)\delta(r,1) & \delta(s',1)\delta(r',0) \\
        + \delta(s,1)\delta(r,0) & \delta(s',0)\delta(r',1)
    \end{aligned}
\end{equation}
Following the expression of Kronecker deltas as a polynomial in the space of possible states $\mathcal{S}$, these expressions amount to joint moments of the distribution of the pairs of interacting states before and after the interaction.

Note that the distribution of a finite number of discrete states has a finite number of independent moments. If each transition is associated a different interaction, then the selector functions will essentially constrain all the joint moments of the joint distribution of states before and after the transition. This defines the distribution uniquely without the need for maximum caliber, which may be somewhat disappointing. 

However, not all transitions must be associated a different interaction (although it can be this way if the application requires it), and the intuition gained by interpretation of constraining these moments as selector functions is still valuable. Moreover, this points to a relation to methods of moment closure applied to study dynamic networks\cite{demirel2014moment,kuehn2016moment,wuyts2022mean}, which has already been explored in literature\cite{rogers2011maximum}. Finally, note that if the states are discrete but infinite, or continuous, specifying any finite number of moments will not suffice to define a unique distribution.

\section{The choice of a link}
\label{app:links}

While for the processes considered in this paper a link is chosen uniformly throughout those of a fixed network at each time, the method of maximum caliber allows us to consider different cases. First, it is clear that just as we chose the conditional probability $\rho(ij|\vec{s}) = P_G(i,j)$ to use in the update of the node states $\rho(\vec{s'}, \vec{s}) = \sum_{ij} \rho(\vec{s'}|ij, \vec{s})P_G(i,j) \rho(\vec{s})$ at each time, we could consider an explicit time dependence in the network and have some evolving distribution $P_{G(t)}(i,j)$ with no change save that the distribution is different at each time. However, this assumes that the evolution of the network is independent of the states.

More generally, we may be interested in a choice of a link that depends on the states of nodes. For example, consider the problem of the mean cover time of a random walk on a network. This can be captured by imagining a network on which node states $\mathcal{S} =\{0,1,2\}$ represent an uncovered node, one that has already been covered, and one that is occupied by a walker respectively. The interaction table for this process can be chosen, for example, to be \cref{tab:cover_time_interactions}, representing that the walker always moves from its current node to the one it interacts with (regardless of whether the new node was already covered), that node states change only if the pair involves a node occupied by a walker, and that the number of walkers does not change through the transitions. 
\begin{table}[ht!]
    \centering
    \begin{tabular}{c|ccccccccc}
          & 00 & 01 & 02 & 10 & 11 & 12 & 20 & 21 & 22 \\
    \hline
    00    & 1  &    &    &    &    &    &    &    &    \\
    %\hline
    01    &    & 1  &    &    &    &    &    &    &    \\ 
    %\hline
    02    &    &    &    &    &    &    &    &    &    \\
    %\hline
    10    &    &    &    & 1  &    &    &    &    &    \\
    %\hline
    11    &    &    &    &    & 1  &    &    &    &    \\
    %\hline
    12    &    &    &    &    &    &    & 1  & 1  &    \\
    %\hline
    20    &    &    &    &    &    &    &    &    &    \\
    %\hline
    21    &    &    & 1  &    &    & 1  &    &    &    \\
    %\hline
    22    &    &    &    &    &    &    &    &    & 1  
    \end{tabular}
    \caption{Transition probabilities of interactions in mean cover time of a network, with nodes being either uncovered ($0$), covered ($1$), or occupied by a walker ($2$). Empty cells of the table are forbidden transitions.} 
    \label{tab:cover_time_interactions}
\end{table}

Assume now that the initial distribution of states has a single walker. In the perspective of the results developed here, we may simply choose links uniformly in the network, as links that do not involve the walker will remain the same. However, this means that the average steps between changes in the states of the nodes of the entire network will depend on the degree (number of connections) at the node where the walker is located at each time. Thus, it is more natural to choose a link according to which nodes contain a walker. For this, we can consider choosing links according to the distribution
\begin{equation}
    \rho(ij|\vec{s}) = \delta(s_i, 2) \frac{a_{ij}}{K^{out}_i} \, .
\end{equation}
This is essentially a uniform choice among the $K^{out}_i = \sum_j a_{ij}$ links connected from node $i$ when it is occupied by the walker $s_i = 2$.

However, we can no longer factor out the choice of a link when we aggregate over states of the entire network except one node in \cref{eq:consecutive_states_general}. In this particular case, because the choice of a node occupied by a walker $\delta(s_i,2)$ is multiplicative, then we can still factor out a term that depends only on the network structure (in this case $a_{ij} / K^{out}_i$) but for more general distributions $\rho(ij | \vec{s})$ this is not necessarily the case.

Moreover, note that while we have simply postulated the conditional probability $\rho(ij|\vec{s})$, it may be more precise to understand that this too is a distribution obtained by choosing constraints and then finding the maximum entropy distribution. For example, constraints that result in $\rho(ij|\vec{s}) = P_G(ij)$ can be expressed as
\begin{equation}
    \sum_{ij, \vec{s}} \delta(i,k) \delta(j,l)\rho(ij, \vec{s}) = P_G(k,l) ~~~~ \forall ~~ k,l \,  .
\end{equation}
Note that because the constraints do not depend on the state of the nodes, then the distribution will not either, following the same logic as before.

\section{Beyond pairwise interactions}
\label{app:ternary}

The assumption of no self-loops in the network implies that all interactions are pairwise, that is involving two strictly different nodes. This motivates considering transitions between pairs of possible states from our construction of the interaction table. However, note that in a network with self-loops, when a link is chosen between a node and itself then transitions are between one of the possible states and another instead of between a possible pair of states and another.

For this, we can introduce a self-interaction table $A$ analogous to the interaction table $B$, with $S$ columns and rows (where $S$ is the number of states), each corresponding to a state $s$ and $s'$ of a self-interacting node before and after the transition. We then define selector functions $g_\beta(s',s)$ for self-interactions $\beta$ identified between these transitions just as we did for the interaction table, but now depending only on the states of a node before and after the interaction.

For example, for the competition-limited population growth process on a network with self-loops, we might consider a self-interaction table such as \cref{tab:self_interactions} capturing spontaneous births $b'$, $0 \rightarrow 1$, and deaths $d'$, $1 \rightarrow 0$.
\begin{table}[ht!]
    \centering
    \begin{tabular}{c|cc}
        & 0 & 1 \\
    \hline
    0   & n  &  $d'$ \\
    %\hline
    1   & $b'$ & n \\ 
    \end{tabular}
    \caption{Self-interaction table for a node in the competition-limited population growth process} 
    \label{tab:self_interactions}
\end{table}
The selector functions of these self-interactions are then 
\begin{equation}
    \begin{aligned}
        g_{b'}(s',s) &= \delta(s',1)\delta(s,0) \\
        g_{d'}(s',s) &= \delta(s',0)\delta(s,1) \, .
    \end{aligned}
\end{equation}
With the selector functions of the self-interactions $g_\beta$ and those of the pairwise interactions $f_\alpha$, we can then define constraint functions $G_m$ besides those $F_n$ already introduced for the pairwise interactions. The constraints then take the forms

\begin{equation}
    \begin{aligned}
        &G_{m}(\vec{s'},ij,\vec{s}) = \delta(i,j) g_\beta(s'_i, s_i) \\
        &F_n(\vec{s'}, ij, \vec{s}) = (1 - \delta(i,j))f_\alpha(s'_i, s_j', s_i, s_j) \, .
    \end{aligned}
\end{equation}
These constraints can then be seen as evaluating the selector functions, either for the self-interactions or the pairwise ones, depending on whether the link $ij$ is a self-loop or not (respectively).

The effect that the presence of self-loops $P_G(i,i) > 0$ and new constraints $G_\beta(s'_i,s_i)$ have on the resulting dynamics of \cref{eq:single_cell_dynamics} is quite straightforward. First, the interaction table $B(s_i',s_j'|s_i,s_j)$ in \cref{eq:interactions_and_unchanged,eq:new_state_transitions} becomes $B(s_i',s_j'|s_i,s_j)^{1 - \delta(i,j)} A(s_i'|s_i)^{\delta(i,j)}$. This essentially evaluates the transition probabilities according to the pairwise interaction table $B$ or the self-interaction table $A$ depending on whether the link is between different nodes $\delta(i,j) = 0$ or a self-loop $\delta(i,j) = 1$. Note that the product over transitions of non-interacting nodes $\prod_{l \neq i,j} C(s'_l,s_l)$ will simply run over all nodes except $i$ if $j = i$. 

When we introduce the expression into \cref{eq:population_successive_probabilities}, then instead of \cref{eq:population_transition_probabilities} we obtain
\begin{widetext}
\begin{equation}
    \rho(\vec{s'}. \vec{s}) = \sum_i \left\{ P_G(i,i) A(s'_i|s_i) \prod_{l \neq i} \delta(s_l', s_l) + \sum_{j \neq i} P_G(i,j) B(s_i',s_j'|s_i,s_j) \prod_{l \neq i,j} \delta(s_l',s_l) \right\} \rho(\vec{s}) \, .
\end{equation}
\end{widetext}
From here on, marginalisation to the states $s_k',s_k$ of a particular node $k$ in the network follows the same logic as before, namely calculating $\rho_k(s',s) = \sum_{\vec{s'}, \vec{s}} \delta(s_k',s') \delta(s_k, s) \rho(\vec{s'}|\vec{s})\rho(\vec{s})$. The main difference is that now, besides contributions from pairs of nodes $i,j$, both different $i \neq k \neq j$ or one equal and one different $i = k \neq j$ or $i \neq k = j$, from the node of interest $k$, we also have contributions from a single node $i$ that can be either equal $i = k$ or different $i \neq k$ to $k$. Note that the probability of all links (i.e. either self-loop or not) is normalised $\sum_{i,j} P_G(i,j) = 1$, not the distribution of self-loops or pairwise links alone.

Rather similarly, we might be interested in considering interactions between three nodes. For example, in a predator prey system, we might give the possibility from prey $s_i = 1$ at a node $i$ to run away from a predator $s_j = 2$ at a node $j \neq i$ it interacts with if a neighbouring node $j \neq k \neq i$ is empty $s_k = 0$. We must then consider a ternary interaction table with $S^3$ rows and columns, each representing a triplet of states of the nodes that interact. We will not represent this interaction table due to its size (already $27 \times 27$ for the three states of the predator-prey system), but the identification of interactions among the possible transitions is the same as for the self-interaction and pairwise interaction tables.

The choice of interacting nodes in this case is somewhat more subtle. To choose three nodes, we need to choose two links connecting them. But not any two links can be chosen, as an arbitrary pair of links can involve up to four nodes. Moreover, if the two links we choose are the same, this can be interpreted as pairwise interactions taking place instead of ternary ones, just as self-interactions are captured by a special type of link between nodes. In general, we can imagine selecting a triplet of nodes by first selecting one link in the network and then another from one end of the connection $\rho(i,j,k) = \rho(k|i,j) \rho(i,j)$. These selections may depend on the states of nodes or not with the same caveats described in the previous subsection, and may be the same probabilities for different links as in the cases here or not. 

We must then take constraints that depend on three nodes in general, where all of them being the same one indicates a self-interaction, only two different nodes and one equal to one of the others a pairwise one, and involving three different nodes a ternary interaction. These can be distinguished by constraints involving different products of Kronecker deltas $f_A(i,j,k) = \delta(i,j) \delta(j,k)$, $f_B(i,j,k) = \delta(i,j) (1 - \delta(j,k) - \delta(i,k))$, and $f_C(i,j,k) = (1 - \delta(i,j))(1 - \delta(j,k))(1 - \delta(i,k))$ respectively. The transition matrix $B$ in \cref{eq:population_successive_probabilities} then becomes $A^{f_A(i,j,k)}B^{f_B(i,j,k)} C^{f_C(i,j,k)}$, where $C$ is the transition probabilities of the ternary interaction tables. The sum in the same equation must be carried out over all possible triplets of nodes, which will evaluate the appropriate interaction table according to which nodes are different and which are the same.

\end{document}